\documentclass[]{aa} 
\usepackage{graphicx,amssymb,amsmath,multirow}
\usepackage[usenames,dvipsnames]{xcolor}
\usepackage{rotating}
\usepackage{txfonts}
\usepackage{lscape}
\usepackage{amsmath}
\usepackage{natbib,twoopt}
\usepackage{longtable}
\usepackage{multicol}
\bibpunct{(}{)}{;}{a}{}{,} 
\newcommandtwoopt{\citeyearads}[3][][]%
{\href{http://adsabs.harvard.edu/abs/#3}{\citeyear[#1][#2]{#3}}}
\usepackage[breaklinks,colorlinks,linkcolor=red,citecolor=blue,urlcolor=blue]{hyperref}

\newcommand{\feh}{[Fe/H]}
\newcommand{\kms}{km s$^{-1}$}
\newcommand{\teff}{$\mathrm{T}_{\mathrm{eff}}$}
\newcommand{\logg}{$\log g$}
\newcommand{\vmic}{$v_{\mathrm{mic}}$}

\newcommand{\vsini}{$v{\sin{i}}$}

\newcommand{\epsFor}{\object{$\epsilon$~For}}

\newcommand{\psiPhe}{\object{$\psi$~Phe}}

\newcommand{\fig}[1]{Fig.~\ref{#1}}
\newcommand{\sect}[1]{Sect.~\ref{#1}}
\newcommand{\FeI}{$\ion{Fe}{i}$}
\newcommand{\FeII}{$\ion{Fe}{ii}$}
\newcommand{\angdi}{$\theta_{\mathrm{LD}}$}
\newcommand{\Fbol}{F$_{\mathrm{bol}}$}
\newcommand{\cand}{\textit{candidate}}

\defcitealias{Heiter2015}{Paper I}

\newcommand{\gaia}{{\it Gaia}}

\begin{document}

\title{Gaia FGK benchmark stars: new candidates at low-metallicities \thanks{Based on data obtained from the ESO Science Archive Facility under request number pdjofre132105 }  }

\author{
	K. Hawkins \inst{\ref{ioa}}
	\and P. Jofr\'e \inst{\ref{ioa}}
	\and U. Heiter \inst{\ref{uppsala}} 
	\and C. Soubiran \inst{\ref{LAB}}
	\and S. Blanco-Cuaresma \inst{\ref{Geneva}}
	\and L. Casagrande  \inst{\ref{ANU}}
	\and G.~Gilmore \inst{\ref{ioa}} 
	\and K.~Lind \inst{\ref{MPIA}}
	\and L.~Magrini \inst{\ref{Arcetri}}
	\and T.~Masseron \inst{\ref{ioa}}
	\and E.~Pancino \inst{ \ref{Arcetri}, \ref{bol1},\ref{ASI}}
	\and S.~Randich \inst{\ref{Arcetri}}
	\and C.~C. Worley \inst{\ref{ioa}}
	}
	%

\offprints{ \\ 
K. Hawkins, \email{khawkins@ast.cam.ac.uk}}

\institute{
	Institute of Astronomy, University of Cambridge, Madingley Road, Cambridge CB3 0HA, United Kingdom \label{ioa}
	\and Department of Physics and Astronomy,  Uppsala University, Box 516, 75120 Uppsala, Sweden \label{uppsala}
	\and Universit\'e de Bordeaux - CNRS, LAB - UMR 5804, BP 89, 33270, Floirac, France \label{LAB} 
	\and {Observatoire de Gen\`eve, Universit\'e de Gen\`eve, CH-1290 Versoix, Switzerland} \label{Geneva}
	\and{Research School of Astronomy and Astrophysics, Mount Stromlo Observatory, The Australian National University, ACT 2611, Australia} \label{ANU}
	\and{Max Planck Institute for Astronomy Konigstuhl 17, D-69117 Heidelberg, Germany}\label{MPIA}
	\and {INAF/Osservatorio Astrofisico di Arcetri, Largo Enrico Fermi 5 50125 Firenze, Italy}\label{Arcetri} 
	\and INAF/Osservatorio Astronomico di Bologna, Via Ranzani 1, 40127 Bologna, Italy  \label{bol1} 
	\and ASI Science Data Center, Via del Politecnico snc, 00133, Roma, Italy \label{ASI}
	}

\authorrunning{Hawkins et al. }
\titlerunning{Gaia metal-poor benchmark stars}
   \date{}

\abstract{We have entered an era of large spectroscopic surveys in which we can measure, through automated pipelines, the atmospheric parameters and chemical abundances for large numbers of stars. Calibrating these survey pipelines using a set of "benchmark stars" in order to evaluate the accuracy and precision of the provided parameters and abundances is of utmost importance. The recent proposed set of Gaia FGK benchmark stars of \cite{Heiter2015} has up to five metal-poor stars but no recommended stars within $-2.0 <$ \feh $< -1.0$ dex. However, this metallicity regime is critical to calibrate properly.}{ In this paper, we aim to add {\it candidate} \gaia\ benchmark stars inside of this metal-poor gap. We began with a sample of 21 metal-poor stars which was reduced to 10 stars by requiring accurate photometry and parallaxes, and high-resolution archival spectra.}{The procedure used to determine the stellar parameters was similar to \cite{Heiter2015} and \cite{Jofre2014} for consistency. The difference was to homogeneously determine the angular diameter and effective temperature (\teff) of all of our stars using the Infrared Flux Method utilizing multi-band photometry. The surface gravity (\logg) was determined through fitting stellar evolutionary tracks. The \feh\ was determined using four different spectroscopic methods fixing the \teff\ and \logg\ from the values determined independent of spectroscopy.}{We discuss, star-by-star, the quality of each parameter including how it compares to literature, how it compares to a spectroscopic run where all parameters are free, and whether \FeI\ ionisation-excitation balance is achieved.}{From the 10 stars, we recommend a sample of five new metal-poor benchmark candidate stars which have consistent \teff, \logg, and \feh\ determined through several means. These stars, which are within $-1.3 <$ \feh\ $< -1.0$, can be used for calibration and validation purpose of stellar parameter and abundance pipelines and should be of highest priority for future interferometric studies.} 
 
 \maketitle

\section{Introduction}

Chemodynamical studies of our Galaxy are beginning to use large samples of stars as a result of in multi-object spectroscopic surveys (e.g. Gaia-ESO, APOGEE, GALAH, and others). In particular, the recently launched \gaia\ satellite will undoubtedly revolutionise our understanding of the Milky Way with accurate parallaxes and proper motions, and accompanying spectral information for more than a billion stars. Combining data from the many multi-object spectroscopic surveys which are already underway, and the rich dataset from \gaia\ will be the way forward in order to disentangle the full chemo-dynamical history of our Galaxy.  One example is the Gaia-ESO Public Spectroscopic Survey \citep[GES,][]{Gilmore2012,Randich2013}, which aims to provide atmospheric parameters and elemental abundances of more than 10$^5$ stars. Another example is the Australian GALAH survey \citep{De_silva2015}, which will undoubtedly contain large numbers of metal-poor stars, and the Apache Point Galactic Evolution Experiment (APOGEE) survey \citep{Eisenstein2011}, which samples giant stars across a broad range in metallicity. In the future, even larger datasets will be produced, such as the southern 4MOST survey \citep{de_jong2012} or its complimentary northern survey WEAVE \citep{Dalton2014}.  

Our methods to do stellar spectroscopy, in particular, to determine the main atmospheric parameters including effective temperature (\teff), surface gravity (\logg) and metallicity (\feh), have necessarily evolved towards a more automatic and efficient way. However, these methods need to be calibrated in order to judge their performance. This calibration can be properly done with a set of well-known stars, or {\it benchmark stars}. In addition, the multiple surveys need to be corrected for systematic offsets between them in order to compare results. This work is about the assessment of such stars. 

Beside astrometry, Gaia will produce, for most stars, atmospheric parameters of stars through a pipeline named APSIS \citep{Bailer-Jones2013}. For the calibration of APSIS, we have, in previous reports on this subject, defined a set of stars that cover different parts on the Hertzsprung-Russell diagram (HRD) in the FGK spectral range \citep[henceforth Paper I,][]{Heiter2015}. We attempted to cover a wide range in metallicities, such that these stars would represent a large portion of the \gaia\ observations.  We have called this sample the \gaia\ FGK benchmark stars (GBS, Paper I). The \teff\ and \logg\ of the current set of GBS have been determined with fundamental relations, independently from spectroscopy, making use of the star's angular diameter (\angdi) and bolometric flux (\Fbol) combined with its distance (Paper I). The metallicity is then determined by using a homogeneous library of spectra. That library is described in \citep[][henceforth Paper II]{2014A&A...566A..98B}. This library is analysed to determine the metallicity based on the adopted values for \teff\ and \logg\ \citep[][henceforth Paper III]{Jofre2014}. High spectral resolution analyses not only yield atmospheric parameters but also individual abundances, thus the same library has been used to derive the abundance of 4 alpha elements and 6 iron-peak elements \citep[][henceforth Paper IV]{Jofre2015}.

These stars have been shown to be an excellent sample to calibrate the stellar parameter determination pipelines of the Gaia-ESO Survey \citep[][Recio-Blanco et al. in prep]{Smiljanic2014} or other spectroscopic surveys and studies \citep[e.g.][]{2014MNRAS.443..698S, 2014arXiv1409.2258D, 2014arXiv1409.7703L,De_silva2015, Boeche2015, Hawkins2016}. However, the calibrations are currently limited by less than a handful of metal-poor main-sequence stars in our initial GBS sample \citep[e.g. see the calibration paper by][]{Smiljanic2014}. The reason is that metal-poor stars are normally further away and thus fainter, making it impossible to measure their \angdi\ accurately with current interferometric instruments except in very rare cases. The metallicity regime around \feh\ $\sim -1.0$ dex is particularly important because this represents the transition between several Galactic components \citep[e.g.][]{Venn2004, Nissen2010, Bensby2014, Hawkins2015b}. For example, the halo is thought to have a mean metallicity of --1.5 with a dispersion of 0.50~dex and the thick disk has a mean metallicity of --0.50~dex with a dispersion of 0.25~dex. Thus at \feh\ $\sim$--1.0 dex, the thick disk and halo components are entangled. Therefore, it is critical to calibrate this metallicity regime correctly. 
 
 \begin{figure}
\includegraphics[width=\columnwidth]{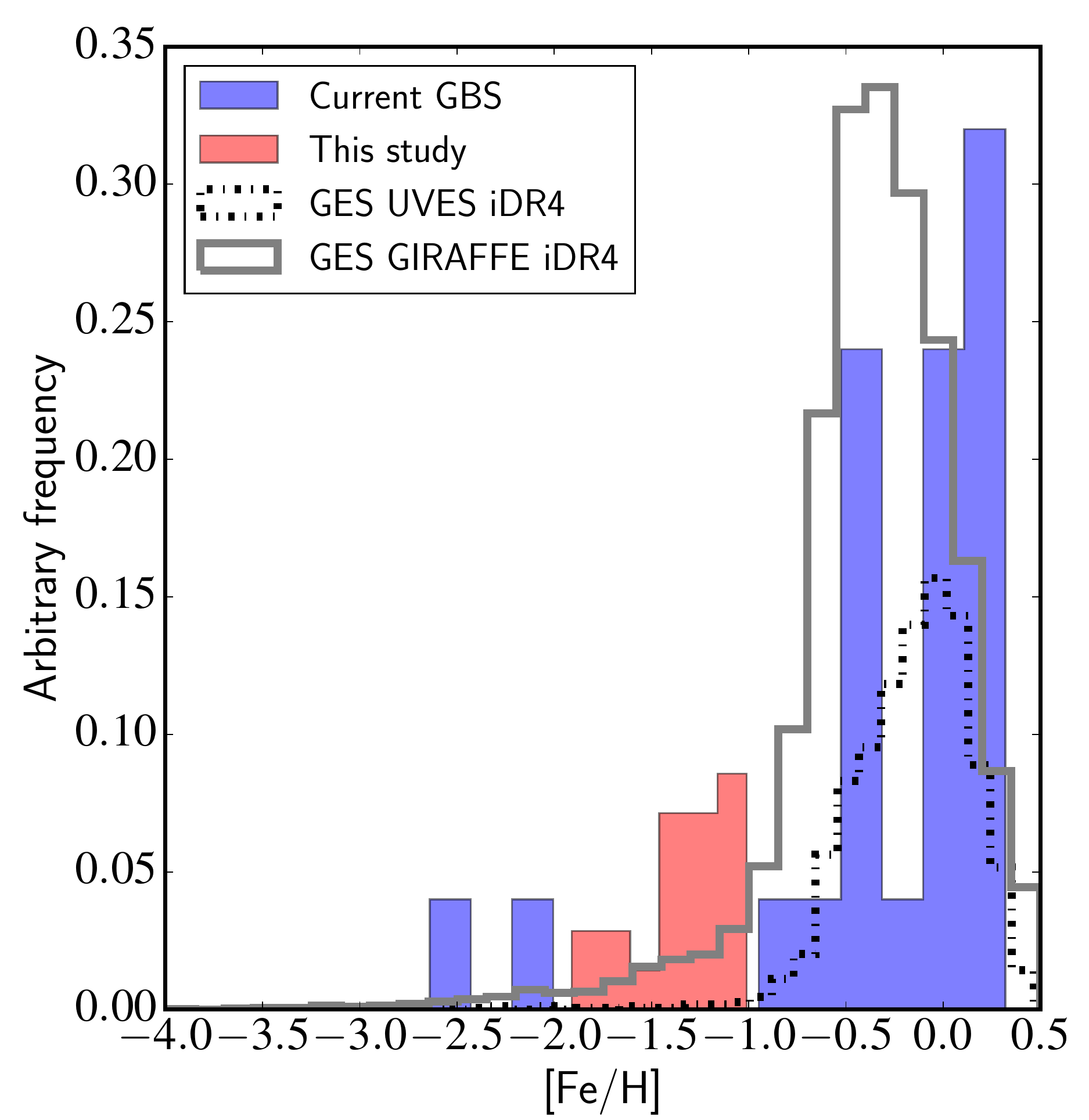}
  \caption{The \feh\ distribution of the current GBS sample from Paper III (blue filled histogram) and the selected sample of metal-poor GBS candidates (red filled histogram). The GES iDR4 metallicity distribution from the UVES sample and GIRAFFE sample are shown as a black dash-dotted histogram and gray solid histogram, respectively. }
  \label{fig:metdist}
\end{figure}

Among the set of current GBS, nearly 20\% (6 stars) have radius and bolometric flux estimated indirectly using photometric relations. At least one of the two current (recommended) metal-poor GBS have radius and bolometric flux estimated indirectly using photometric relations. In this paper, we use similar and consistent relations to include more metal-poor stars in a homogeneous way. We do this because for many of these GBS candidates \angdi\ can not yet be reliably measured with interferometry. In particular, systematic effects might still be the major limitation at the sub-milliarcsec level \citep[e.g.][]{Casagrande2014}. 

In the current set of GBS there are a total of five metal-poor stars with \feh\ $<$ --1.0 dex (\psiPhe, HD122563, HD84937, HD140283, Gmb 1890). However among these five, three have not been recommended for calibration or validation purposes in Paper I.  HD140283 was not recommended because of the large uncertainties in the \teff\ which is likely a result of a calibrated bolometric flux which had large systematic differences between the photometric and spectroscopic values. Gmb 1830 has a highly uncertain \teff\ which could be due to calibration errors in the interferometry (Paper I) and thus it was not recommended. The measured angular diameter of \cite{Creevey2015} yields an effective temperature that is more than 400 K lower than the spectroscopic \teff. Additionally, the cool M giant star \psiPhe\ was not recommended, in part, because of an uncertain metallicity caused by the inability of the methods employed to properly deal with the molecular features which heavily crowd the spectrum. 

This leaves only two metal-poor stars which have metallicities below --2.0 dex and effectively no stars with $-2.0 <$ \feh\ $< -1.0$ dex. We aim to provide a set of new GBS {\it candidate} stars inside of the metal-poor gap listed above. These new stars ultimately will allow the astronomical community and spectroscopic surveys to extend their calibrations based on the benchmark stars possibly reaching into the critical regime of  $-1.3 <$ \feh\ $< -1.0$ dex. The metallicity distribution of the recommended set for calibration and validation purposes from Paper I (blue histogram) and the additional metal-poor candidate stars (red histogram) are shown in Fig. \ref{fig:metdist}. In the background of that figure is the metallicity distribution of the full recommended sample of stars from the GES iDR4 UVES (black dash-dotted histogram) and GIRAFFE (gray solid histogram) spectra \citep[for more information on UVES see][]{Dekker2000}. A sizable fraction of the stars in the GES iDR4 are in the metal poor regime and thus a proper calibration through metal poor GBS is necessary. The recommended stellar parameters (and metallicity) of the GES iDR4 stars have been determined by spectral analysis of several methods (nodes) whose results have been homogenized and combined (details of this will be published in Hourihane et al. 2016 in prep).  

In this fifth work of the series, we define a new set of GBS \cand\ stars inside of the metal-poor gap. We note that these candidates do not have \angdi\ measurements and should remain as candidates until an \angdi\ can be measured directly, at least for a handful, in the near future. In addition, we aim to provide a set of metal-poor stars with predicted \angdi\ which can be used as the input for future interferometric studies. 

As such, this paper is organised in the following way: in Sect. \ref{sec:sample+method} we begin by selecting a sample of relatively bright metal-poor stars that have archival spectra. We then describe the several methods that we have used to determine the \teff\ (Sect. \ref{subsec: teff}) and log g (Sect. \ref{subsec: logg}). Fixing these parameters, we determined the metallicity using methods consistent with Paper III, which we describe in Sect. \ref{subsec:feh}. In Sect. \ref{sec:results} we present the results of the parameter analysis and discuss, star-by-star, the quality of the parameters and recommend a new set of metal-poor benchmark stars. We also compare our results with what is known about these stars in the literature. Finally, in Sect. \ref{sec:conclusion} we summarize our analysis and recommendations.

\section{Sample} \label{sec:sample+method}

The initial target list was selected using the PASTEL database requiring the following: (1) 4500 $<$ \teff\ $<$ 6500~K, (2) --2.0 $<$ \feh\ $<$ --1.0 dex, and (3) there were at least four \teff\ and metallicity estimates in the literature, since 1990, with a standard deviation of less than 100 K and 0.1 dex, respectively. The third criterion was used filter out stars where there are obvious discrepancies in the stellar parameters or the star was ill-behaved in order to maximise the chance that after our analysis, the stars will have metallicities and parameters in the regime of interest. These criteria result a total of 21 stars including Gbm1380 (HD103095). The metallicity distribution of these 21 stars can be found as the red histogram in Fig. \ref{fig:metdist}. We further required there to be known BVJHK photometry with defined uncertainties less than 0.15 mag in  order to compute accurate photometric \teff. This criterion reduced the sample to 17 stars, removing BD+053640, HD199289, HD134440. We also required there to be a known, and non-negative, parallax with a relative uncertainty better than 50\%. This criterion removed four stars (HD206739, HD204543, HD063791, HD083212). Finally, we required spectra in the ESO and NARVAL archives. This last criterion removed four stars (HD21581\footnote{This star now has spectra available in the ESO archive but was not public when the target selection for this project was completed}, HD023439A/B\footnote{This is a spectroscopic binary system in which neither component had an ESO/NARVAL spectrum.}). The stars BD+053640, HD206739, HD063791, and HD083212 also do not have high-resolution spectra in the ESO/NARVAL archives. 
 
 This reduced the sample to the final version of 10 selected stars. We note that after the above cuts, we have mostly selected stars with --1.3 $<$ \feh $<$ --1.0 dex, which is highly appropriate given that the interface of the thick disk, accreted halo, and possibly even the thin disk is within the regime \citep[e.g.][]{Nissen2010, Bensby2014, Hawkins2015b} and there is a lack of such stars in the current GBS. Throughout the rest of this paper, we will consider and discuss only these 10 stars.

In Table \ref{tab:generalinfo}, we present the basic collected information for the new candidates including the sky position (J2000 right ascension, RA, and declination, DEC) and the mean and dispersion of the stellar parameters taken from PASTEL. Additionally, the photometric and parallax information can be found in Table \ref{tab:photinfo}. Their $B$- and $V$-band photometry were taken from the General Catalogue of Photometric Data \citep[henceforth GCPD,][]{Mermilliod1997}. Where the $B$- and $V$-band photometry was not defined the in the GCPD catalogue the Simbad database was used. The $J_{\mathrm{2MASS}}$ and $K_{\mathrm{2MASS}}$ magnitudes are sourced from the 2MASS catalogue \citep{Cutri2003}. The adopted reddening values, E$(B-V)$ were taken from \cite{Melendez2010}, \cite{Casagrande2010}, and \cite{Casagrande2011}. The parallax for each star was adopted from the updated analysis of the Hipparcos catalogue \citep{VanLeeuwen2007}. The tables have been separated by those stars which have been selected for further analysis and those which have not for clarity.
 
The final sample that we focus on in this paper contains those 10 metal-poor stars, all covering  the metallicity regime that we are most interested in, namely metal-poor (\feh\ $\sim$ --1.0 dex) stars with an emphasis on dwarf stars. More than half of these stars were suggested in Appendix B of Paper I. The analysis presented here is consistent with the previous papers in the series (Paper I, Paper II, Paper III) allowing the parameters of these metal-poor stars to be added to the GBS sample covering a wide and well sampled parameter space in the HRD.\\ 

 \begin{table*}
\caption{General information on metal-poor benchmark candidates.} 
\begin{tabular}{c c c c c c c c c c c c } 
\hline\hline 
Star&RA&DEC&\teff&$\sigma$\teff&N&log g&$\sigma$log g&N&[Fe/H]&$\sigma$[Fe/H]&N\\
 & (J2000) & (J2000) & (K) & (K) & & (dex) & (dex) & & (dex) & (dex) &\\
 \hline
 & & & & & Selected & & & & \\
 \hline
BD+264251&21:43:57.12&+27:23:24.00&5991&97&8&4.30&0.36&7&--1.27&0.08&7\\
HD102200&11:45:34.24&--46:03:46.39&6119&52&10&4.22&0.16&7&--1.22&0.06&7\\
HD106038&12:12:01.37&+13:15:40.62&6012&68&9&4.36&0.09&4&--1.31&0.04&4\\
HD126681&14:27:24.91&--18:24:40.44&5567&84&21&4.59&0.17&12&--1.18&0.09&12\\
HD175305&18:47:06.44&+74:43:31.45&5085&58&15&2.49&0.25&13&--1.43&0.07&14\\
HD196892&20:40:49.38&--18:47:33.28&5954&94&9&4.16&0.24&8&--1.03&0.08&9\\
HD201891&21:11:59.03&+17:43:39.89&5883&68&35&4.33&0.15&28&--1.05&0.08&28\\
HD218857&23:11:24.60&--16:15:04.02&5119&40&7&2.50&0.34&6&--1.91&0.09&7\\
HD241253&05:09:56.96&+05:33:26.75&5879&94&13&4.35&0.15&9&--1.06&0.06&9\\
HD298986&10:17:14.88&--52:29:18.71&6177&82&8&4.23&0.06&5&--1.33&0.04&5\\
\hline
 & & & & & Not Selected & & & & \\
 \hline
BD+053640&18:12:21.88&+05:24:04.41&5051&83&9&4.59&0.15&5&--1.20&0.10&5\\
HD021581&03:28:54.48&--00:25:03.11&4889&61&11&2.15&0.20&6&--1.67&0.08&7\\
HD023439A&03:47:02.12&+41:25:38.12&5059&73&13&4.51&0.11&13&--1.06&0.08&13\\
HD023439B&03:47:02.63&+41:25:42.56&4808&70&6&4.55&0.09&6&--1.04&0.09&6\\
HD063791&07:54:28.72&+62:08:10.76&4715&73&9&1.75&0.07&9&--1.68&0.08&9\\
HD083212&09:36:19.95&--20:53:14.75&4512&55&15&1.37&0.32&13&--1.46&0.06&13\\
HD103095&11:52:58.76&+37:43:07.23&5071&76&44&4.65&0.17&36&--1.34&0.10&38\\
HD134440&15:10:12.96&--16:27:46.51&4817&82&16&4.61&0.11&11&--1.44&0.08&11\\
HD199289&20:58:08.52&--48:12:13.45&5895&60&12&4.36&0.23&9&--1.01&0.06&10\\
HD204543&21:29:28.21&--03:30:55.37&4667&66&15&1.30&0.22&11&--1.80&0.10&12\\
HD206739&21:44:23.94&--11:46:22.84&4662&33&8&1.70&0.00&5&--1.58&0.02&5\\
\hline \hline
\end{tabular}
\\ 
\tablefoot{The stellar parameters for each star were compiled using the PASTEL database \citep{Soubiran2010}. The \teff, $\sigma$\teff, log g, $\sigma$log g, \feh, and $\sigma$\feh\ represent the mean and dispersion of the stellar parameters from N references in the PASTEL database. }
\label{tab:generalinfo}
\end{table*}

As in Paper I-IV, we chose stars that have been widely studied in the past. Table \ref{tab:generalinfo} indicates there are between 4 -- 35 studies for each star. However, as seen below, these studies are very different from each other (using different procedures to determine the stellar parameters) and thus the advantage of this work is to homogenise the stellar parameters with respect to Paper I-IV so that they can be ingested into the current GBS. 

The parameters given in Table \ref{tab:generalinfo} have been determined through a variety of means. For example, the \teff\ has been determined through both photometric \citep[e.g.][]{Alonso1996, Nissen2002, Ramirez2005, Jonsell2005, Masana2006, Reddy2006, Gonzalez-Hernandez2009, Casagrande2010, Casagrande2011, Ishigaki2012} and spectroscopic \citep[e.g.][]{Gratton1996, Nissen1997, Gratton2000, Mishenia2000, Fulbright2000, Gratton2003, Sousa2011} means. In some cases the spectroscopic \teff\ is determined by fitting the wing of the strong Balmer H features, usually H$\alpha$ or H$\beta$ \citep[e.g.][]{Axer1994, Mashonkina2000, Gehren2004}. Since the distance is known, the \logg\ is largely derived using the parallax \citep[e.g.][]{Gratton2000, Gehren2004, Jonsell2005}. However, in some cases the Fe ionisation balance \citep{Axer1994, Fulbright2000, Sousa2011, Ishigaki2012} or Mg-triplet wing fitting \citep[e.g.][]{Mashonkina2000} has been used. Metallicity is determined from the analysis of iron lines under 1D-LTE approximations in most of the works \citep[e.g.][]{Axer1994, Fulbright2000, Jonsell2005, Valenti2005, Sousa2011, Ishigaki2012}. Extensive discussions of these works and our results are found in Sect. \ref{sec:results}.

 \begin{table*}
\caption{Photometry and Parallax of Metal-Poor Benchmark Candidates.} 
\begin{tabular}{c c c c c c c c c c c c c} 
\hline\hline 
Star&$B$&$\sigma B$&$V$&$\sigma V$&N&$J_{\rm 2MASS}$&$\sigma J$&$K_{\mathrm{2MASS}}$&$\sigma K$&$\pi$&$\sigma \pi$&E$(B-V)$\\
& (mag) & (mag) & (mag) & (mag) & & (mag) & (mag) & (mag) & (mag) & (mas) & (mas) & (mag)\\
\hline
& & & & & & Selected & & & & & \\
\hline
BD+264251&10.52&0.03&10.05&0.03&2&8.98&0.02&8.64&0.02&9.03&1.68&0.007\\
HD102200&9.21&0.01&8.76&0.00&2&7.69&0.02&7.38&0.02&13.00&0.98&0.005\\
HD106038&10.63&0.01&10.16&0.01&9&9.11&0.03&8.76&0.02&9.98&1.57&0.003\\
HD126681&9.90&0.01&9.31&0.01&4&8.04&0.02&7.63&0.02&21.04&1.12&0.000\\
HD175305&7.93&0.02&7.17&0.01&6&5.61&0.02&5.06&0.02&6.39&0.36&0.000\\
HD196892&8.73&0.02&8.23&0.02&5&7.18&0.03&6.82&0.02&16.15&0.93&0.000\\
HD201891&7.89&0.01&7.38&0.02&10&6.25&0.02&5.93&0.02&29.10&0.64&0.000\\
HD218857&9.60&0.11&8.88&0.11&3&7.40&0.02&6.87&0.02&3.21&1.09&0.019 \\
HD241253&10.24&0.00&9.72&0.00&2&8.64&0.03&8.29&0.02&8.66&1.77&0.001\\
HD298986&10.46&0.03&10.03&0.03&4&9.04&0.02&8.74&0.02&6.61&1.41&0.004\\
\hline
& & & & & & Not Selected & & & & & \\
\hline
BD+053640&11.16&...&10.43&...&0&8.85&0.04&8.34&0.04&15.58&1.82&...\\
HD021581&9.54&0.01&8.71&0.00&2&6.98&0.02&6.41&0.02&4.03&1.00&...\\
HD023439A&8.93&0.01&8.18&0.02&4&6.62&0.02&6.12&0.02&46.65&2.63&...\\
HD023439B&9.26&0.03&8.77&0.01&4&6.95&0.02&6.35&0.02&50.72&1.02&...\\
HD063791&8.81&0.02&7.90&0.01&0&6.05&0.04&5.43&0.02&1.07&0.73&...\\
HD083212&9.39&0.04&8.32&0.03&3&6.31&0.02&5.61&0.02&0.96&0.77&...\\
HD103095&7.19&0.02&6.44&0.02&2&4.94&0.20&4.37&0.03&109.99&0.41&...\\
HD134440&10.22&0.02&9.43&0.01&1&...&...&...&...&35.14&1.48&...\\
HD199289&8.82&...&8.30&...&1&7.18&0.02&6.84&0.02&18.95&0.76&...\\
HD204543&9.17&0.02&8.30&0.01&0&6.46&0.02&5.78&0.02&--0.13&1.08&...\\
HD206739&9.45&0.03&8.60&0.02&0&6.70&0.02&6.03&0.02&1.93&1.17&...\\
\hline \hline
\end{tabular}
\\ \\
 \tablefoot{The $B$ and $V$ magnitudes were sourced from the GCPD catalogue \citep{Mermilliod1997} with N number of references.  In cases where the $B$ and $V$ were not found in the GCPD catalogue, they were taken from the Simbad database. The $J_{\mathrm{2MASS}}$ and $K_{\mathrm{2MASS}}$ magnitudes are sourced from 2MASS \citep{Cutri2003}. All parallaxes were taken from a reanalysis of the Hipparcos catalogue \citep{VanLeeuwen2007}. The adopted reddening values were taken from \cite{Melendez2010}, \cite{Casagrande2010}, and \cite{Casagrande2011}.}
\label{tab:photinfo}
\end{table*}

\section{Determination of effective temperature}\label{subsec: teff}
\teff\ was determined in two ways: (1) using \angdi-photometric calibrations \citep{VanBelle1999, Kervella2004, Benedetto2005, Boyajian2014} with the Stefan-Boltzmann law, and (2) using the IRFM \citep[e.g.][]{Blackwell1977,Blackwell1979,Blackwell1980,Casagrande2006, Casagrande2010}. In Sect. \ref{subsub:angdi_teff} we describe the first procedure and in Sect. \ref{subsubsec:IRFM} we discuss the second procedure. 

\subsection{Deriving temperature using angular diameter-photometric relationships}
\label{subsub:angdi_teff}
To compute the \teff, we used Equation 1 of Paper I which relates the \teff\ to the bolometric flux, \Fbol, and the \angdi. We estimated the \Fbol\ using the photometric relationship outlined in Equations 8 and 9 of \cite{Alonso1995} which rely on the $V$ and $K$ photometry. We note that the photometric relationship to obtain the \Fbol\ required that the $K$ magnitude was in the Johnson rather than 2MASS bandpasses. Thus, we converted the 2MASS photometry (columns 7 and 9 in Table 2) bands into the  Johnson system using the following relationship:
 \begin{equation}
 K_{\mathrm{J}} = K_{\mathrm{2MASS}} - 0.1277(J-K)_{\mathrm{2MASS}} + 0.0460 ,
 \label{eq:2MASS2J}
 \end{equation}
\\where $ K_{\mathrm{J}}$,  $K_{\mathrm{2MASS}}$ are the $K$-band magnitude in the Johnson and 2MASS systems, respectively. The 2MASS subscript refers to the 2MASS J, K, and (J-K), and the J subscript to Johnson system. This relationship was obtained by combining Equations 6, 7, 13, and 14 from \cite{Alonso1994} and Equations 12 and 14 from \cite{Carpenter2001}\footnote{We note that Equation 12 and 14 from \cite{Carpenter2001} have been updated in 2003. These updates can be found at http://www.astro.caltech.edu/~jmc/2mass/v3/transformations/. The difference between the 2001 and 2003 values is negligible. For example, the mean difference in  $K_{\mathrm{J}}$ is 0.005 mag leading to a change in \teff\ on the order of less than 8~K.}. The uncertainty in $ K_{\mathrm{J}}$ was determined by propagating the uncertainty in the $K_{\mathrm{2MASS}}$ and $(J-K)_{\mathrm{2MASS}}$. We note here that the photometry was corrected for reddening using the values in column 13 of Table \ref{tab:photinfo}. These corrections are very small and have the effect of changing the \angdi\ on the order of less than 1\% and \teff\ by less than 30~K when compared to the raw photometric values.

 \begin{figure}
\includegraphics[width=\columnwidth]{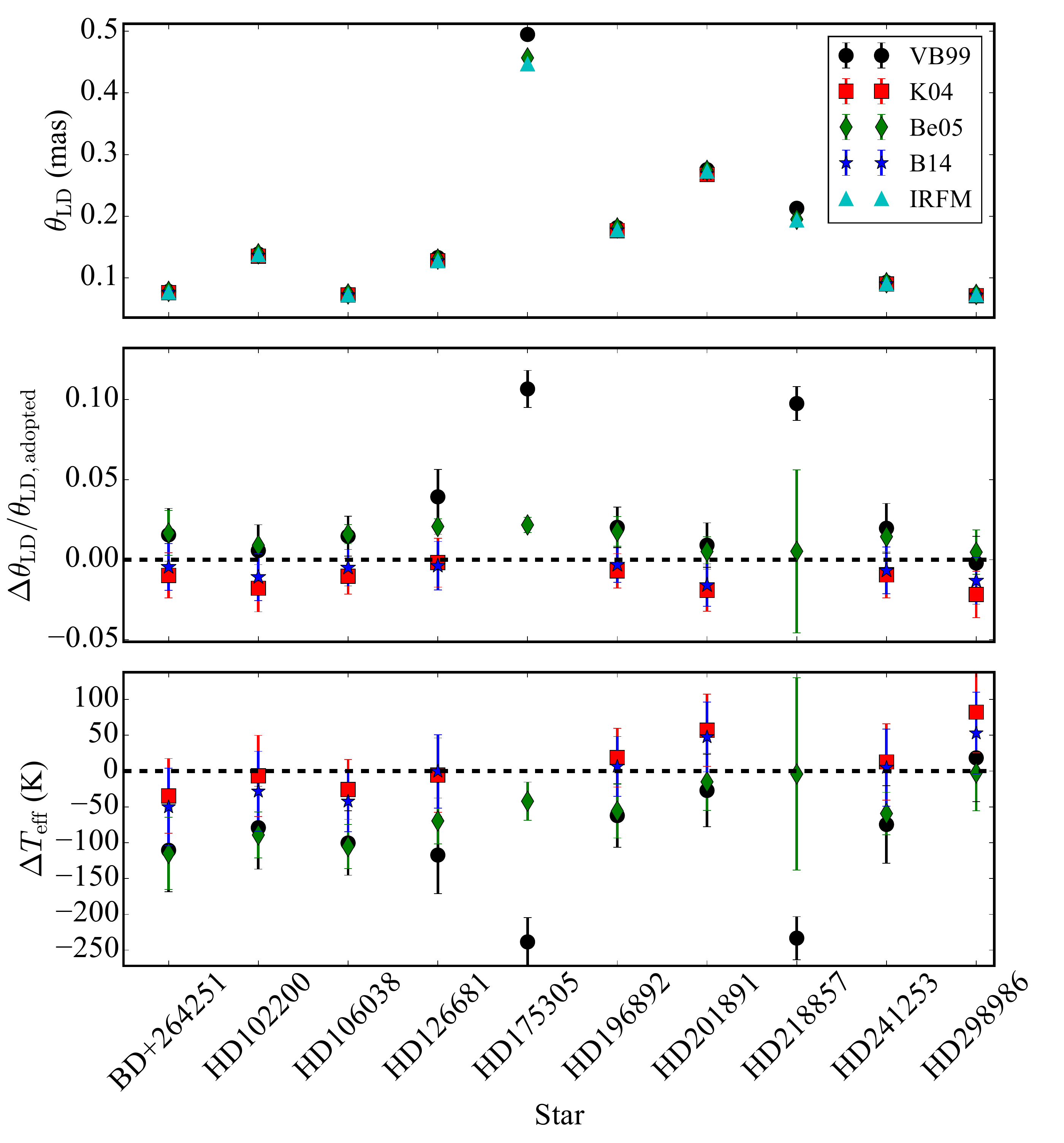}
  \caption{Top Panel: The computed \angdi\ for each star from the four \angdi-photometric relationships:  \cite[K04,][]{Kervella2004} is represented by red squares,  \cite[VB99,][]{VanBelle1999} is represented by black circles, \cite[Be05,][]{Benedetto2005} is represented by green diamonds, \cite[B14,][]{Boyajian2014} is represented by blue stars. In addition, the infrared flux method (IRFM) is also displayed as (cyan triangles). Middle Panel: $\Delta\theta/\theta_{\mathrm{adopted}}$ for each star. Here $\Delta\theta_{\mathrm{LD}} =$ \angdi\ - $\theta_{\mathrm{LD,adopted}}$. The adopted \angdi\ is that computed from the IRFM. Bottom Panel: Comparison of the \teff\ for each star, computed from the \angdi-photometric relationships, with the adopted value from the infrared flux method.} 
  \label{fig:angdi}
\end{figure}

The \angdi\ was determined indirectly through photometric relationships. We have made use of four separate \angdi-photometric relations in order to test the robustness of this procedure \citep{VanBelle1999, Kervella2004, Benedetto2005, Boyajian2014}. The first set of calibrations used were taken from the work of \cite{VanBelle1999}. We determined the angular diameter of all stars by taking the average of the \angdi-($B-V$) relation (their Equation 2) and \angdi-($V-K$) relation (their Equation 3). The second set of calibrations, which was used only for the dwarf stars, were taken from the photometric relationships of \cite{Kervella2004}. Just as above, we averaged the \angdi-($B-V$) relation (their Equation 22) and \angdi-($V-K$) relation (their Equation 23). We note this procedure was used for the \angdi\ for the GBS HD22879 and \epsFor\ in Paper I. The third set of calibrations were from \cite{Benedetto2005}. We computed the \angdi\ of all stars using the \angdi-($B-V$) relation (their Equations 1 and 2). The final set of calibrations used were from \cite{Boyajian2014}. We made use of their \angdi-($B-V$) relation (their Equation 4) which is only applicable to the dwarf stars.  

The results of the \angdi\ and \teff\ computed using the various \angdi-photometric calibrations above can be found in Fig. \ref{fig:angdi}. In top panel of Fig. \ref{fig:angdi}, we compare the \angdi, in miliarcseconds (mas), of each star and relation used. We also plot the \angdi\  computed from the infrared flux method \cite[hereafter IRFM, see Sect.  \ref{subsubsec:IRFM} and][for more details]{Casagrande2006,Casagrande2010,Casagrande2014}. In the middle panel of Fig. \ref{fig:angdi} we show the relative difference between the four \angdi-photometric relations with that computed from the IRFM. In the bottom panel of Fig. \ref{fig:angdi}, we compare the \teff\ derived from the different \angdi-photometric calibrations and that computed from the IRFM. In most cases the \angdi\ from each of the photometric calibrations are consistent (within 1$\sigma$) with each other and the \angdi\ from the IRFM. 

As noted by Paper I, we choose to use the (V-K)-\angdi\ relationships because they have the smallest dispersion in the fitted relationship (on the order of less than 1\%) compared to other photometric colours. These equations are created by relating the \angdi\ of dwarf, subgiant, and giant stars determined via interferometry to their $(V-K)_{\mathrm{J}}$ colour and $ K_{\mathrm{J}}$ magnitude \citep[e.g.][]{VanBelle1999, Kervella2004, Benedetto2005, Boyajian2014}. While it is likely that the brightest star (HD175305) may soon have direct \angdi\ measurements, most of these stars are dim, making direct interferometric \angdi\ measurements difficult with current instruments. Thus for the moment, we have the only option to rely on the photometric calibrations for \angdi\ and \Fbol. 

It is important to note that recent studies \citep[e.g.][]{Creevey2012, Creevey2015} have indicated the \angdi-photometric relationship may underestimate the \angdi\ particularly at low metallicities. This is likely because the \angdi-photometric relationship are often only constrained by less than a handful, around 2--3, metal-poor stars \citep[e.g. see Fig. 5 of][]{Kervella2004}. Since the \teff\ is proportional to \angdi$^{-0.5}$,  underestimating the \angdi\ causes the \teff\ to be overestimated. We also made use of the IRFM because it has the advantage of including not only information from $V$ and $K$ but also a broad range of photometry improving the \teff\ estimate and predicted \angdi\ (see Sect. \ref{subsubsec:IRFM}). 

We are also prompted to use the IRFM because there is a relatively large disagreement (on the order of 10\% which causes differences in \teff\ of more than 300 K) between the \angdi\ of the giant stars in our sample using the calibrations of \cite{VanBelle1999} and \cite{Benedetto2005}. The reason for this discrepancy is currently not clear. One explanation is that there are intrinsic errors in the procedures that were used to determine the fitted relations. For example, reddening was not taken into account when relating the photometric colours to \angdi\ in the work of \cite{VanBelle1999} which in part could cause discrepancies in their fitted relationships. In addition, it is important to note that a weakness of using these relations is that they do not include dependencies on \feh. As a result, many of these relations perform best around solar metallicity, by construction. 

\subsection{Infrared Flux Method}
\label{subsubsec:IRFM}

In addition to the $\theta_{\rm{LD}}$-photometric relationships used in the previous section, we also made use of the infrared flux method (IRFM). This is one of the least model-dependent techniques to determine effective temperatures in stars, and it was originally devised to obtain stellar angular diameters with an accuracy of a few percent \citep{Blackwell1977, Blackwell1979, Blackwell1980}. Our analysis is based on the implementation described in \cite{Casagrande2006, Casagrande2010}. 

The basic idea is to recover for each star its bolometric flux and infrared monochromatic flux, both measured on the Earth. Their ratio is then compared to that obtained from the two same quantities defined on a surface element of the star, i.e., the bolometric flux $\sigma T_{\mathrm{eff}}^4$ and the theoretical infrared monochromatic flux. The only unknown parameter in this comparison is \teff, which can be obtained (often with an iterative scheme, as described further below). For stars roughly earlier than M-type, the theoretical  monochromatic flux is relatively easy to compute because the near infrared region is largely dominated by the continuum, with a nearly linear dependence on \teff\ (Rayleigh-Jeans regime) and is largely unaffected by other stellar parameters such as metallicity and surface gravity. This minimizes any dependence on model atmospheres, and makes the IRFM complementary to most spectroscopic methods, where instead \teff\ is often degenerate with gravity and metallicity. Once the bolometric flux and the effective temperature are known, the limb-darkened angular diameter is self-consistently obtained from the IRFM. Since most of the times fluxes are derived from multi-band photometry, the problem is ultimately reduced to a derivation of fluxes in physical units, i.e. it depends on the photometric absolute calibration. Without exaggeration, this is the most critical point
when implementing the IRFM, since it sets the zero-point of the \teff\ scale. In our case, the absolute calibration has been anchored using solar twins, and the zero-point of the resulting effective temperature scale thoroughly tested \citep{Casagrande2010, Casagrande2014, Datson2012, Datson2014}.

For the sake of this work, the bolometric flux was recovered using multi-band photometry (Johnson-Cousins $BV(RI)_C$ and 2MASS $J_{\mathrm{2MASS}}$, $H_{\mathrm{2MASS}}$, $K_{\mathrm{2MASS}}$) and the flux outside of these bands estimated using theoretical model fluxes from \cite{Castelli2004}. For each star [Fe/H] and \logg\ were fixed to the GBS recommended values. Whereas an iterative procedure was adopted in \teff, starting with an initial guess, and iterating the IRFM until convergence within 1~K was reached. Despite all candidate GBS being relatively nearby, some of them might be slightly affected by extinction. When available, we adopted the reddening values derived from interstellar Na I D lines \citep{Melendez2010} or from \cite{Casagrande2010} or \cite{Casagrande2011} for the remaining cases. 

\begin{figure}
\includegraphics[width=\columnwidth]{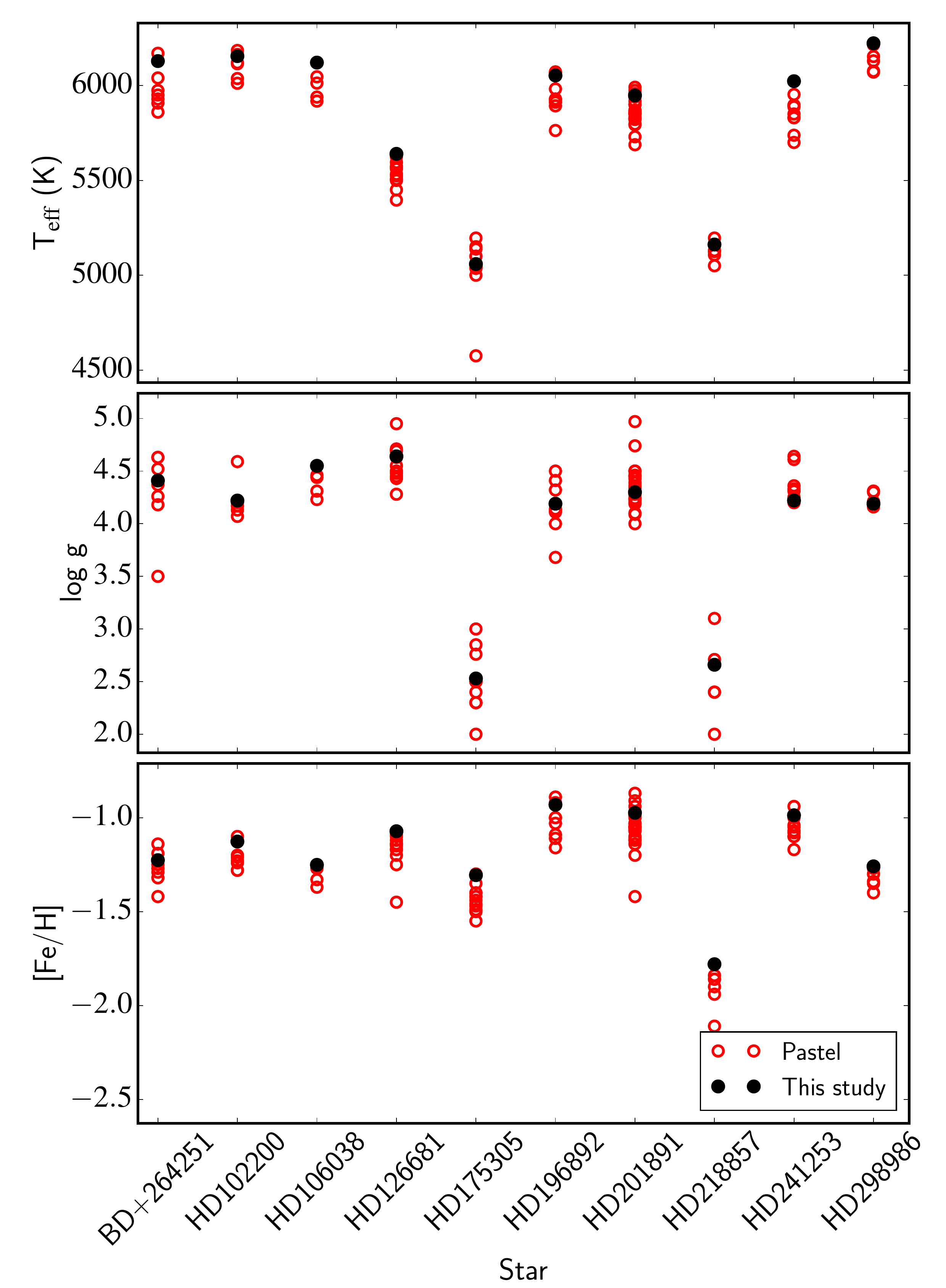}
\caption{The adopted \teff\ (top panel), log g (middle panel), and \feh\ (bottom panel) of the metal-poor GBS candidate stars (black closed circles) compared with the values from the literature (open red circles) sourced from the PASTEL catalogue.}
\label{fig:final_lit}
\end{figure}

Ultimately, we adopted the \teff\ computed from the IRFM, as opposed to the \angdi-photometric calibrations, in large part because it provides a robust estimate of the \angdi\ for the two problematic giant stars making use of the available full broad band photometry rather than the ($V-K$) colour. In addition, in this way we have \angdi\ and \teff\ from both giant and dwarf stars that are computed using a homogenous framework. We note that for all dwarf stars, except for BD+264251, the IRFM temperature agrees very well with all four \angdi-photometric relationships described in Sect. \ref{subsub:angdi_teff} within the 1$\sigma$ uncertainty. This is also the case for the giant stars when considering the calibration of \cite{Benedetto2005} but not that of \cite{VanBelle1999}. The top panel of Fig. \ref{fig:final_lit} shows the adopted \teff\ from the IRFM with respect to literature values obtained from the PASTEL database. It indicates that the \teff\ determined from the IRFM are systematically larger, by $\sim$60~K, compared to the mean \teff\ from the PASTEL database. However, the IRFM is favored in this study over the PASTEL database because it is internally consistent. The reason for this minor discrepancy is unclear. The adopted \Fbol, \angdi, and \teff and their uncertainties determined via the IRFM  can be found in Table \ref{tab:computedinfo}. 


 \section{Determination of surface gravity}\label{subsec: logg}
The surface gravity was determined using the same procedure as in Paper I. We briefly summarize this method below. The \logg\ was determined using the adopted relationship g~=~GM~/~R$^2$ where G is Newton's gravitational constant, M is the mass of the star and R is its radius. The radius of the star was estimated using the adopted \angdi\ which is listed in Col. 1 of Table \ref{tab:computedinfo} and the parallax listed in Col. 11 of Table \ref{tab:photinfo}. The mass for each star was computed by fitting the stellar parameters to a set of stellar evolutionary tracks. In this case, those of Yonsei-Yale\footnote{http://www.astro.yale.edu/demarque/yystar.html} were used \citep[Y$^2$,][]{Yi2003,Demarque2004}.  The fitting procedure is described in Paper I. The luminosity was computed from the bolometric flux and parallax. The \teff\ used was the value adopted from the IRFM. Additionally, the input metallicity was initially assumed to be the mean value from the PASTEL database. The \logg\ does not change significantly when using the final metallicity values described in Sect. \ref{sec:results}. More details about the specific inputs to the Y$^2$ models, comparison of the masses determined from other stellar evolutionary tracks \citep[e.g. Padova,][]{Bertelli2008,Bertelli2009}, and a comparison of the \logg\ determined by this method and others can be found in Sects. 4 and 5 of Paper I. The middle panel of Fig.~\ref{fig:final_lit} indicates that the \logg\ values determined in the this way are consistent with the literature values from the PASTEL database. The grid of stellar models were interpolated with respect to mass and metallicity. The mass was then determined by minimizing the difference between the interpolated models and the position of the star on the HRD. The main source of uncertainty in the \logg\ determined tends to be from the radius (and thus \angdi) compared to the mass (see Sect. 4.1 and Appendix A of Paper I). 

\begin{figure}[t]
      \resizebox{\hsize}{!}{\includegraphics[trim=50 50 30 50,clip]{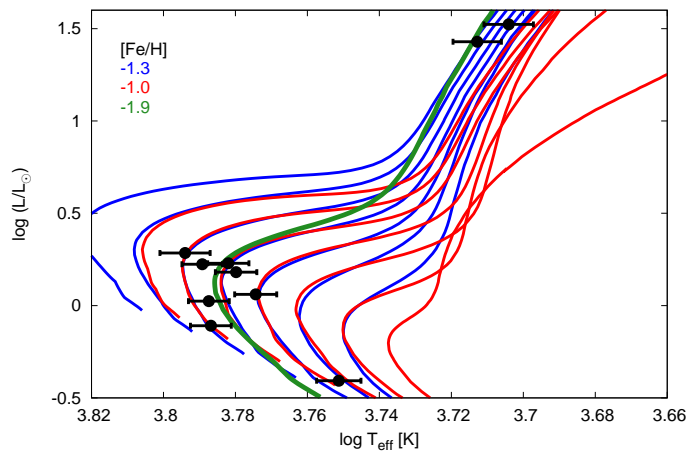}}
  \caption{HRD showing the stellar parameters of the metal-poor benchmark candidates (black circles with error bars) superimposed on \emph{Yonsei-Yale} evolutionary tracks for [$\alpha$/Fe]=+0.3, and [Fe/H]=$-1.3$ (blue lines) and [Fe/H]=$-1.0$ (red lines). Tracks are shown for masses increasing from 0.60~$M_\odot$ to the right to 0.90~$M_\odot$ to the left, in steps of 0.05~$M_\odot$. The green line shows a track for [Fe/H]=$-1.9$ and a mass of 0.70~$M_\odot$, corresponding to the properties of the giant star HD~218857.}
   \label{fig:HRD}
\end{figure}

Figure~\ref{fig:HRD} shows the locations of all stars in the HRD, together with \emph{Yonsei-Yale} evolutionary tracks for different metallicities. Most of the stars cluster around the tracks for 0.8~$M_\odot$, with two dwarfs and the most metal-poor giant at somewhat lower masses. The mass difference of successive tracks (0.05~$M_\odot$) corresponds to the typical uncertainty in mass.

 \begin{table*}
\caption{Adopted parameters for metal-poor benchmark candidates.} 
\begin{tabular}{c c c c c c c c c c c c  } 
\hline\hline 
Star&\angdi&$\sigma$\angdi&\Fbol &$\sigma$\Fbol&\teff&$\sigma$\teff&log g&$\sigma$log g&\vmic&\vsini&$\sigma$\vsini\\
& (mas) &(mas)&(10$^{-11}$ Wm$^{-2}$) &(10$^{-11}$ Wm$^{-2}$)& (K) &(K) &(dex)&(dex)&(\kms)&(\kms)&(\kms)\\
\hline\hline
*BD+264251&0.077&0.001&0.2759&0.0059&6129&80& 4.41&0.16&1.40&2.81&2.10\\
HD102200&0.138&0.002&0.9062&0.0041&6155&80& 4.22&0.07&1.43&1.90&4.59\\
HD106038&0.073&0.001&0.2477&0.0026&6121&80& 4.55&0.14&1.39&0.00&0.00\\
*HD126681& 0.128&0.002&0.5543&0.0050&5640&80& 4.64&0.07&1.16&0.00&0.00\\
HD175305& 0.447&0.006&4.3520&0.6752&5059&80& 2.53&0.14&1.54&5.01&1.92\\
*HD196892&0.178&0.002&1.4130&0.0233&6053&80& 4.19&0.06&1.36&0.00&0.00\\
HD201891&0.273&0.004&3.1154&0.0517&5948&80& 4.30&0.04&1.29&2.93&5.31\\
*HD218857&0.194&0.003&0.8841&0.1625&5162&80& 2.66&0.32&1.58&3.14&6.80\\
*HD241253& 0.091&0.001&0.3642&0.0013&6023&80& 4.22&0.18&1.34&1.89&4.50\\
HD298986&0.073&0.001&0.2691&0.0067&6223&100&4.19&0.19&1.48&4.07&5.98\\
\hline \hline
\end{tabular}
\\ \\
 \tablefoot{The \angdi\ were computed as a part of the IRFM. In addition, the \teff\  and \Fbol\ in this table represents the adopted \teff\  and bolometric flux from the IRFM, respectively. We estimated $v_{mic}$ using the GES relationship of Bergemann and Hill. The uncertainty in $v_{mic}$ was conservatively assumed to be 0.20 \kms\ for all stars. Stars with an asterisk (*) in Col. 1 are currently not recommended (see Sect. \ref{subsec:starbystar} for a star by star discussion on the recommendations.)}
\label{tab:computedinfo}
\end{table*}

\section{Determination of metallicity} \label{subsec:feh}
To determine the metallicity of the candidates, we analysed their spectra. Because these stars were selected from the PASTEL database, they have been previously studied and thus their spectra can be found in archives (see Table \ref{tab:spectra}). Nine stars have previously been observed in the U580 setup of the UVES instrument and the spectra were downloaded from the ESO archives\footnote{http://archive.eso.org/cms.html}. Additionally, one star (HD175305) comes from the archive of the NARVAL spectrograph operated by the T\'elescope Bernard Lyot\footnote{http://tblegacy.bagn.obs-mip.fr/}. The spectra were prepared in the same way as Paper II: they were normalised, corrected by radial velocity and convolved to the lowest common resolution (R = 40000), in the same fashion as in the rest of the GBS. Note the resolution is lower in this case compared to our previous study because we could not find the whole data set with higher resolution. In all cases, the signal to noise ratio (SNR) of the spectra is better than 100 pixel$^{-1}$. The spectra for these stars have been included in the GBS high-resolution spectral library\footnote{ http://www.blancocuaresma.com/s/benchmarkstars/} and can be publicly accessed \citep[for more details on the library consult][]{2014A&A...566A..98B}.

 \begin{table}
  \setlength{\tabcolsep}{3pt}
\caption{Spectra used for this study.} 
\begin{tabular}{c c c c c c} 
\hline\hline 
Star&I&Date$_{\mathrm{obs}}$ & SNR & $R_{\mathrm{in}}$ & Program ID\\
& & &(pixel$^{-1}$) &  & \\
\hline\hline
BD+264251&U&2003-08-09&286&45254&71.B-0529(A)\\
HD102200&U&2001-03-06&160&51690&67.D-0086(A)\\
HD106038&U&2004-03-28&254&45254&072.B-0585(A)\\
HD126681&U&2000-04-09&240&51690&65.L-0507(A)\\
HD175305&N&2010-03-16&150&80000&...\\
HD196892&U&2005-10-15&268&45990&076.B-0055(A)\\
HD201891&U&2012-10-18&107&66320&090.B-0605(A)\\
HD218857&U&2001-10-09&102&56990&68.D-0546(A)\\
HD241253&U&2005-10-08&194&56990&076.B-0133(A)\\
HD298986&U&2000-04-09&173&51690&65.L-0507(A)\\
\hline \hline
\end{tabular}
\\ \\
 \tablefoot{The I, or instrument, is either the U580 setting for the UVES instrument on the Very Large Telescope (denoted by U) or the NARVAL instrument (denoted by N). We note that while the input resolution ($R_{\mathrm{in}}$ = $\lambda/\Delta\lambda$) varies depending on the instrument and setup, we convolved all spectra to a common value of R = 40000. In addition, all of the spectra have a spectral coverage of at least 4760 -- 6840~\AA.}
\label{tab:spectra}
\end{table}

The analysis was done as in Sects. 4.1, 4.2, and 4.3 of Paper III, namely we used several codes. In addition, we used common input material (spectra, atomic data for the line list, \FeI, \FeII\ lines, etc.) and fixed the \teff\ and \logg\ to their adopted values determined in \sect{subsec: teff} and \ref{subsec: logg}, respectively. We made use of the 1D-LTE MARCS atmosphere models \citep{2008A&A...486..951G} and a common set of pre-defined iron lines, which were selected from the ``golden lines'' for metal-poor stars of Paper III. We considered the lines used for HD140283, HD122563, HD84937,  HD22879 and Gmb~1830. Then, by visual inspection, we ensured thatthese lines were present and unblended, in the the spectra of the new candidate stars, obtaining a final list of 131 \FeI\ and \FeII\ lines (see Table 4 from Paper III for the input atomic data). Individual lines used for each star can be found in Tables A1-A10 of the online material. For clarity and reproducibility, in appendix A, we outline the format of the online material. 

In this work, we employed four methods, or nodes, to determine the metallicity. Two methods use the equivalent width (EW) technique which include: (1) Bologna --  based on GALA developed by \cite{2013ApJ...766...78M} and (2) EPINARBO -- based on FAMA developed by \citet{2013A&A...558A..38M}. Both of these methods measure the EWs of individual iron features using the DOOp code \citep{Cantat-Gaudin2014} which is an automated wrapper for the DAOSPEC code \citep{Stetson2008}. The other two methods use spectral synthesis including: (1) BACCHUS/ULB -- developed by T. Masseron \citep{Masseron2006} which made use of the Turbospectrum synthesis code \citep{Alvarez1998, Plez2012} and (2) iSpec -- developed by \citet{2014A&A...569A.111B}. For more details on these methods we refer the reader to Sect. 4.3 of Paper III, Sect. 3.3, Table 4 of Paper IV and the development papers cited above. The first three methods were also employed in our previous metallicity determination in Paper III, while all four methods were used to determine abundances of several elements for the GBS sample (Paper IV). 

The initial metallicity for the analysis was considered to be $\mathrm{[Fe/H]} = -1.00$ dex for all stars. The macroturbulence parameter, $v_{mac}$, was determined simultaneously with the iron abundance, in the same way as in Paper III. The microturbulence parameter, \vmic, was set to the value determined by the GES \vmic\ relationship \citep[e.g.][Paper III, Bergemann et al., in prep]{Smiljanic2014}. 
    
We conducted a total of eight runs which included: the ``main run" fixing the \teff\ and \logg\ and \vmic\ to their adopted values and six ``error" runs where these three fixed values were varied by their $\pm1$$\sigma$ uncertainties listed in Table \ref{tab:computedinfo}. This was done to evaluate the impact of the 1$\sigma$ uncertainty in the adopted parameters on the \feh. In addition, each node solved for the stellar parameters independently using its own procedure, in what we define as the ``free run" or eighth run. We note that in the free run we do not require the different nodes to use the same procedure (e.g. $\sigma$-clipping outlying Fe lines, tolerances of conversion, line selection etc.). This test was done primarily to see how each node performed when not using fixed \teff\ and \logg\ parameters. We emphasize that some of these nodes, particularly the EW nodes, often require a much larger number of lines for best performance. Thus, we remind the reader that the results of the free analysis simply allow us to quantify, in a different way, the benefit of fixing the \teff\ and \logg. We refer the reader to Paper III for a extensive discussion on this matter. 

\begin{figure*}
\includegraphics[width=\textwidth]{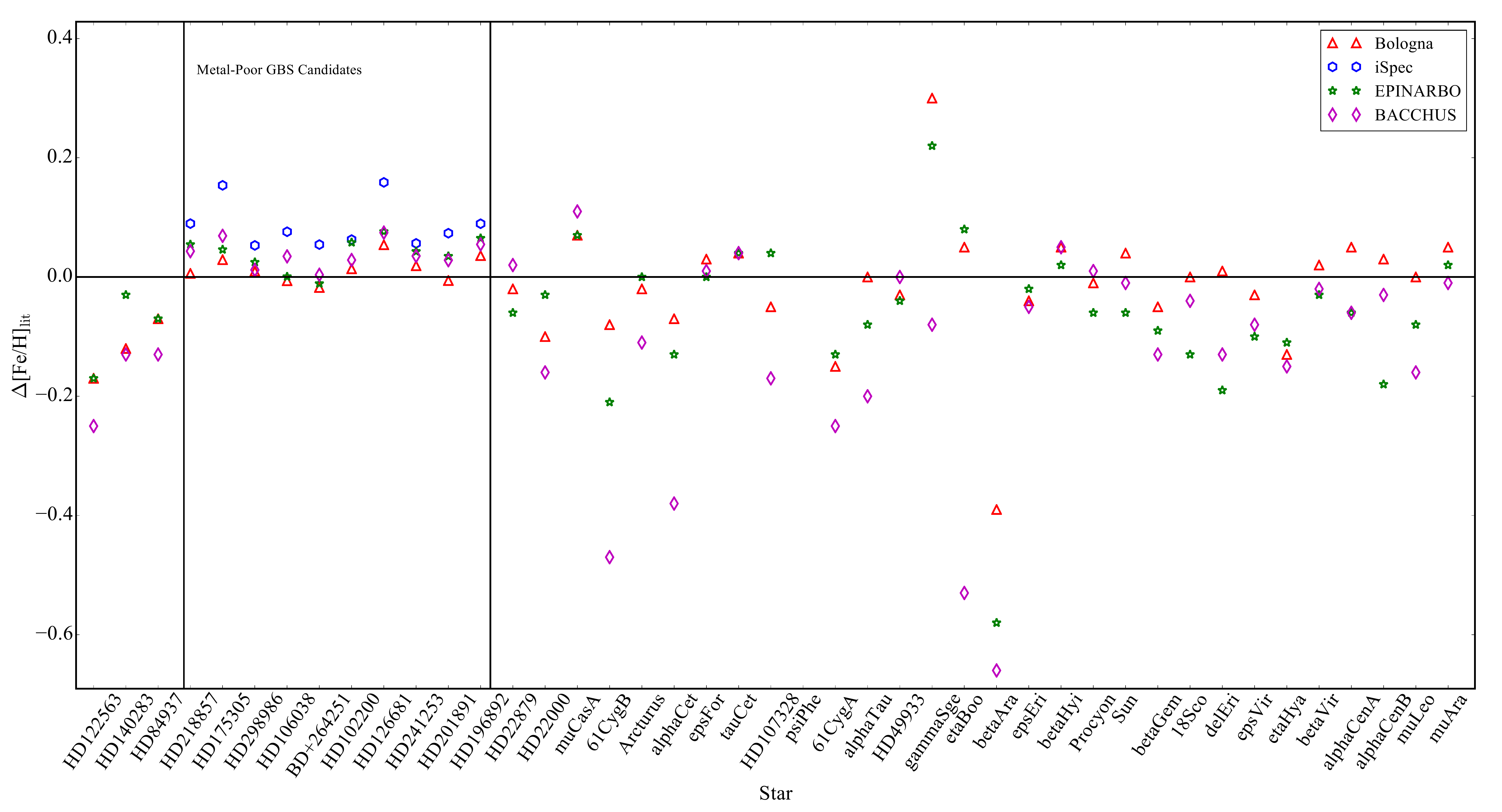}
\caption{The $\Delta$\feh$_{\mathrm{lit}}$ = \feh - \feh$_\mathrm{lit}$ for each star and node for our GBS candidates and the current GBS stars ordered by metallicity. The node symbols are as follows: (1) iSpec is represented as a blue open hexagon, (2) BACCHUS/ULB is represented as a magenta open diamond, (3) EPINARBO is represented as a green open star, and (4) Bologna is represented as a red open triangle. We do not display the nodes that are used in Paper III and not this work. The typical dispersion between the methods is on the order of $\pm$0.03 dex while the typical offset between the literature and the each method is on the order of +0.04 dex.}
\label{fig:nodecomp}
\end{figure*}

A node-to-node comparison of the \feh\ can be found in Fig. \ref{fig:nodecomp} where we plot the metallicity of each star (including the results for the GBS in Paper III) obtained by each node relative to the mean literature value from PASTEL database. We also sort the stars on the x-axis towards increasing metallicity. The y-axis of the figure is the $\Delta$\feh$_{\mathrm{lit}}$, which is defined as \feh --\feh$_\mathrm{lit}$), where \feh\ is the metallicity of the star determined by a specific node and \feh$_\mathrm{lit}$ is the mean \feh\ from the PASTEL database. The name of the star is indicated on the bottom of the figure. We note that only 3 nodes (ULB, Bologna, and EPINARBO) of Paper III were included in this figure. These nodes are the same as in this work. Fig. \ref{fig:nodecomp} indicates that the metallicities from the different nodes for the metal-poor candidates have a standard deviation of 0.028. In addition, the values generally agree well with the literature with a mean offset of +0.04 dex. This is consistent with the offset (+0.04 dex) and standard deviation (0.07 dex) of the FG dwarfs among the GBS (Paper III). The typical node-to-node scatter for the candidate stars are comparable to the GBS in the same \teff\ regime. Again we note that the node abundances for each star were determined by averaging the abundances of each line.  

 \begin{table*}
\caption{Adopted \feh\ for metal-poor benchmark candidates.} 
\begin{tabular}{c c c c c c c c c c c c c  } 
\hline\hline 

Star & [Fe/H] & $\sigma$FeI & $\Delta$(\teff) & $\Delta$(\logg) &  $\Delta$(\vmic) & $\Delta$(LTE) & $\Delta (ion)$ & $\sigma$FeII & N$_{\mathrm{FeI}}$ &  N$_{\mathrm{FeII}}$\\
\hline
*BD+264251  &  -1.23  &   0.07  &   0.09  &   0.02  &   0.02  &   0.03  &  -0.05  &   0.05  &  63  &   8   \\
 HD102200  &  -1.12  &   0.07  &   0.08  &   0.01  &   0.01  &   0.04  &   0.02  &   0.07  &  58  &   8   \\
 HD106038  &  -1.25  &   0.08  &   0.08  &   0.01  &   0.01  &   0.02  &  -0.03  &   0.05  &  66  &   7   \\
 *HD126681  &  -1.07  &   0.06  &   0.05  &   0.01  &   0.01  &   0.01  &   0.02  &   0.05  &  61  &   7   \\
 HD175305  &  -1.29  &   0.06  &   0.06  &  -0.01  & -0.01  &   0.06  &   0.08  &   0.04  &  56  &   8   \\
 *HD196892  &  -0.93  &   0.05  &   0.05  &   0.01  &   0.01  &   0.04  &   0.03  &   0.05  &  68  &   8   \\
 HD201891  &  -0.97  &   0.06  &   0.06  &  0.00  & 0.00  &   0.03  &   0.07  &   0.02  &  68  &   8   \\
 *HD218857  &  -1.78  &   0.07  &   0.11  &   0.04  &   0.04  &   0.06  &   0.01  &   0.05  &  56  &   8   \\
 *HD241253  &  -0.99  &   0.06  &   0.09  &   0.01  &   0.01  &   0.03  &   0.09  &   0.03  &  66  &   7   \\
 HD298986  &  -1.26  &   0.07  &   0.09  &   0.01  &   0.02  &   0.05  &   0.04  &   0.05  &  66  &   6   \\
 
 \hline\hline \\
 \end{tabular}
 \\
 \tablefoot{The \feh\ is the NLTE-corrected and is the recommended value for each star. The $\Delta$(\teff) is the uncertainty in the \feh\ due to the uncertainty in \teff, $\Delta$(\logg) is the uncertainty in the \feh\ due to the uncertainty in \logg, and $\Delta$(\vmic) is the uncertainty in the \feh\ due to the uncertainty in \vmic. $\Delta$(LTE) is the NLTE-corrected \feh\ minus the LTE \feh. $\Delta (ion)$ = [\FeI\ /H] -- [\FeII\/H].  The line-to-line dispersion of \FeI\ and \FeII\ are $\sigma$\FeII\ and  $\sigma$\FeII, respectively. Finally  N$_{\mathrm{FeI}}$ and  N$_{\mathrm{FeII}}$ are the number of \FeI\  and \FeII\ lines used for the analysis, respectively. Stars with an asterisk (*) in Col. 1 are currently not recommended (see Sect. \ref{subsec:starbystar} for a star-by-star discussion on the recommendations).}
 \label{tab:finalmet}
 \end{table*}
 
NLTE-corrected metallicities for each star can be found in Col. 2 of Table \ref{tab:finalmet}. The uncertainty in [Fe/H] due to the uncertainty in \teff, \logg, and \vmic\ can be found in Cols. 4, 5, and 6, respectively. The difference between the LTE and NLTE-corrected metallicity, $\Delta$(LTE), and the difference between the mean \FeI\  and \FeII\ abundance, $\Delta(ion)$, is found in Col. 7 and 8 respectively.  The line-to-line dispersion of \FeI\ , \FeII\, and the number of \FeI\  and \FeII\ lines used in the analysis are listed in Cols. 3, 10, 11, and 12, respectively. Table \ref{tab:finalmet} indicates that the difference between \FeI\ and \FeII\ can be as high as 0.10 dex in the worst cases. The $\Delta(ion)$ values are smaller than for some of the GBS, e.g. HD 122563, where $\Delta(ion)_\mathrm{HD 122563}$ = --0.19 dex (Paper III). We note here that HD 122563 is more metal-poor, with \feh\ = --2.64, than the stars we consider in this paper. On the other, hand the NLTE corrections, which are on the order of 0.05 dex, are similar to those of the current set of GBS. 

We remind the reader that the final metallicity was computed as a mean of NLTE-corrected Fe lines. The NLTE corrections were computed in the same way as Paper III, namely by interpolating over a grid of NLTE corrections outlined in \cite{Lind2012}. For this calculation, the adopted parameters were used. When the NLTE correction for a given line is not available the median of the NLTE corrections is assumed. This is both reasonable and reliable because the NLTE corrections per line are very similar for a single star \citep[e.g.][]{Bergemann2012}. The NLTE correction range from +0.020 to +0.064 dex.

For each \FeI\ and \FeII\ line, run and star we have four measurements (one for each of the nodes) for the iron abundance, which can be found in the tables online. We note here that the EW measurements for the synthesis methods (ULB/BACCHUS, iSpec) are measured for completeness but are not used to measure the abundances. The Fe abundance for each of the selected ``golden" lines,  and its computed NLTE correction can also be found as part of the online material. A description of this online material can be found in appendix A. 

\section{Results and Discussion} \label{sec:results}
In this section. we discuss, on a star-by-star basis the results of the stellar parameter analysis. We discuss the quality of each parameter for each star, separately. In addition, we describe  the node-to-node variation in the stellar parameters. Finally we compare the adopted stellar parameters with those determined spectroscopically. 

As in Pa per~III and Paper~IV, we selected only the lines that were sufficiently strong to have reliable abundances and sufficiently weak to not saturate, that is, line strength or reduced equivalent width (REW) was in the range of $-6.0 \leq$ REW $\leq -5.0$ where REW = $\log(\mathrm{EW}/\lambda)$. For this selection the adopted equivalent width (EW)  was computed by averaging over the four measurements.  Among the selected lines, we computed the mean of the four Fe abundance measurements and calculated its NLTE correction consistent with Paper~III and references therein.

\begin{figure*}
 \resizebox{\hsize}{!}{\includegraphics[trim=0 0 50 200,clip]{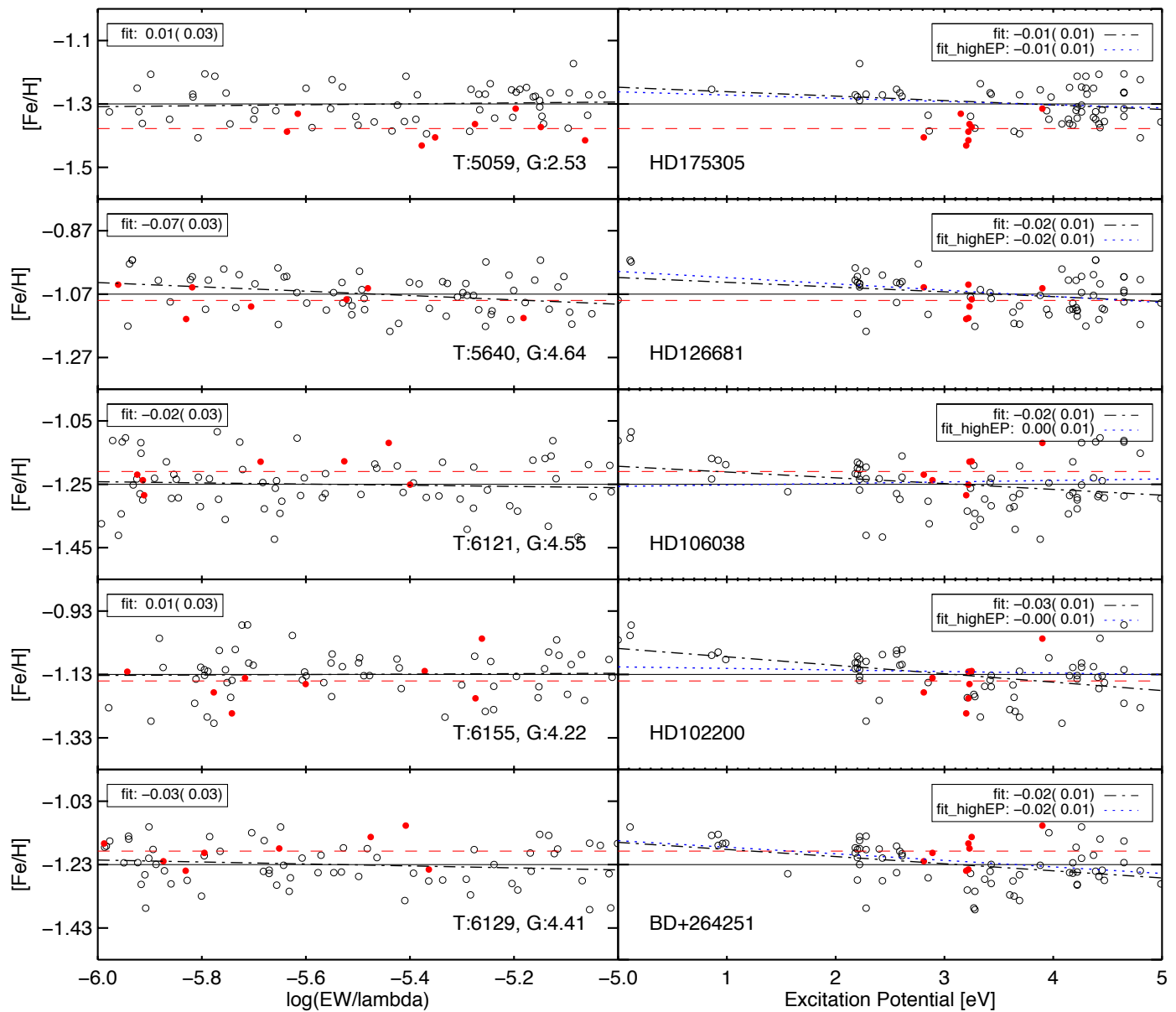}}
  \caption{Final iron abundances as a function of REW (left panels) and EP (right panels) for five stars (BD+264251, HD102200, HD106038, HD126681, HD175305) analysed in this work. Open circles indicate neutral lines while filled red circles indicated ionised lines. A linear regression fit to the neutral lines is performed for all lines (indicated with a black dash-dotted line) and for high-EP lines (EP $\geq 2$eV, indicated by a blue dotted line). The slope of the trend and its standard error are indicated at the top of each panel. A slope is considered to be significant if its absolute value is larger than the standard error. The effective temperature and surface gravity for each star is also indicated at the bottom of the left panels for reference. Dashed red line indicates the mean of the ionised lines. }
  \label{fig:nodes-abu}
\end{figure*}

\begin{figure*}
\resizebox{\hsize}{!}{\includegraphics[trim=0 0 50 220,clip]{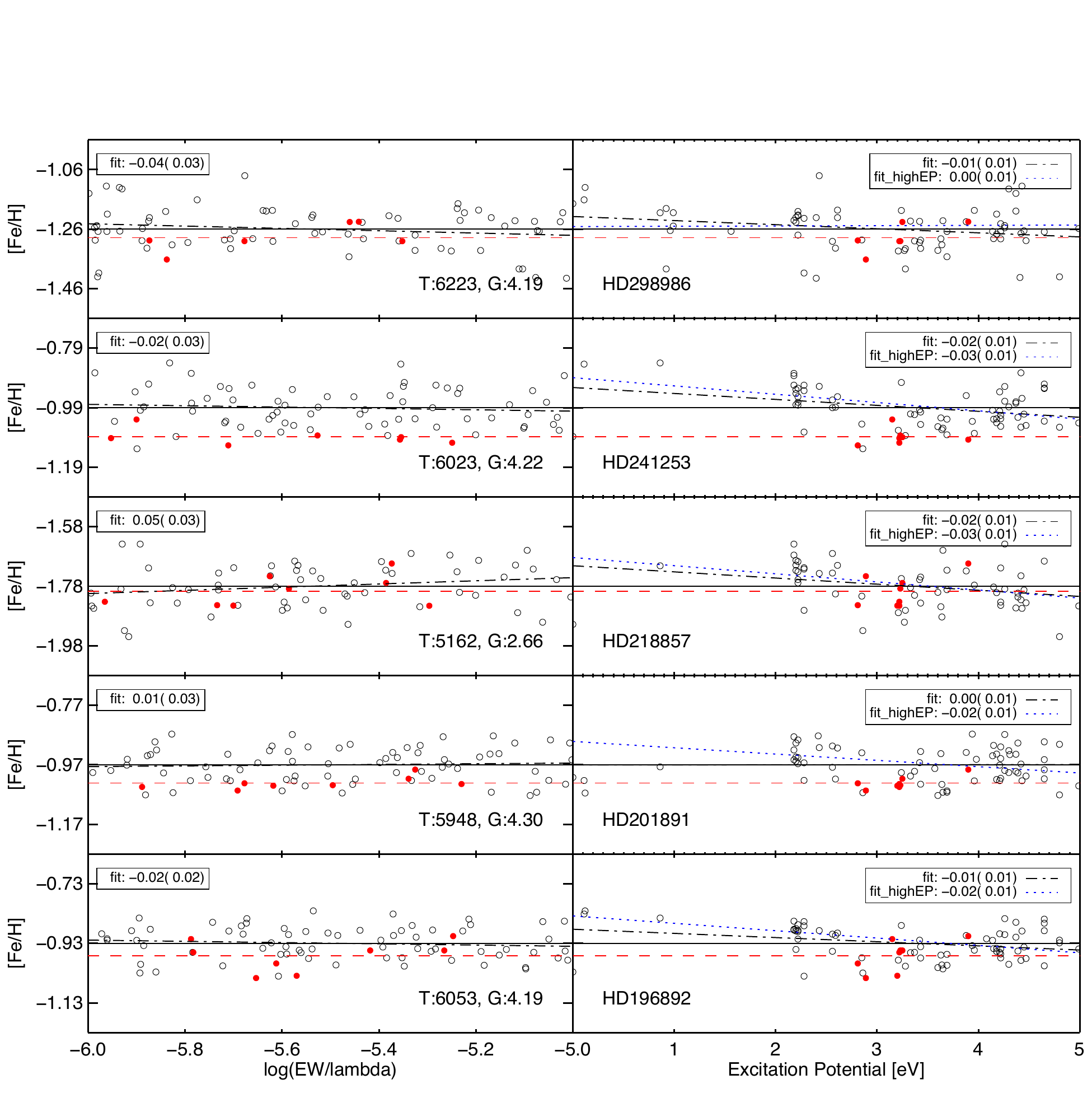}}
  \caption{ Final iron abundances as a function of REW (left panels) and EP (right panels) for the remaining five stars (HD196892, HD201891, HD218857, HD241253, HD298986) analysed in this work. Symbols used are the same as in Fig.~\ref{fig:nodes-abu}.}
  \label{fig:nodes-abu2}
\end{figure*}

To help facilitate the discussion, we plot the final NLTE-corrected abundances for each line and star in \fig{fig:nodes-abu} and \fig{fig:nodes-abu2} using different symbols for neutral and ionised lines. Each star is indicated in a different set of right-left panels. For reference, the star's name is listed in the right panel and its stellar parameters are indicated in the left panel. The left panels show the abundances as a function of REW while the right panels show the abundances as a function of excitation potential (EP).  We performed linear fits to the neutral lines. The slope of the trend and its standard error are indicated at the top of each panel. A slope is considered to be significant if its absolute value is larger than the standard error. We also performed a linear fit to only high EP lines (with EP $\geq 2$eV). We choose this cut because the low-excitation transitions are thought to experience significantly larger departures from 1D, LTE compared to higher excitation transitions \citep[e.g.][]{Bergemann2012}. The red dashed lines correspond to the mean abundances determined from ionised lines.  
 
In Fig. \ref{fig:nodes-abu} and Fig.~\ref{fig:nodes-abu2}, we find that three of the ten stars (HD126681, HD 218857, and HD 298986) have significant trends in REW and Fe abundance indicating an potential issue with their \vmic. Figs.~\ref{fig:nodes-abu} and \ref{fig:nodes-abu2}  also indicates that six of the ten stars have significant trends in the Fe abundance and EP whether using all of the \FeI\  lines or using just the high-EP lines as suggested by \cite{Bergemann2012}.

The criteria for recommending a GBS candidate are as follows: (1) the \teff\ derived from IRFM should be consistent with the \angdi-photometric calibrations, (2) the \teff\ determined via the IRFM and photometric calibrations should be consistent with the spectroscopic \teff\ (i.e. the correlation between EP and Fe abundance should be null), (3) the \logg\ determined via isochrone fitting (assuming the \teff\ from IRFM) should be consistent with the spectroscopic \logg\ (i.e. the mean abundance of \FeI\  should equal that of \FeII). Finally all stars where there is large discrepancies between the recommended parameters and PASTEL (i.e. differences in \teff\ more than 500~K, \logg\ larger than 0.5~dex, \feh\ larger than 0.5~dex) are flagged as suspicious.


\subsection{Star-By-Star Discussion} \label{subsec:starbystar}

 \begin{figure}
\includegraphics[width=\columnwidth]{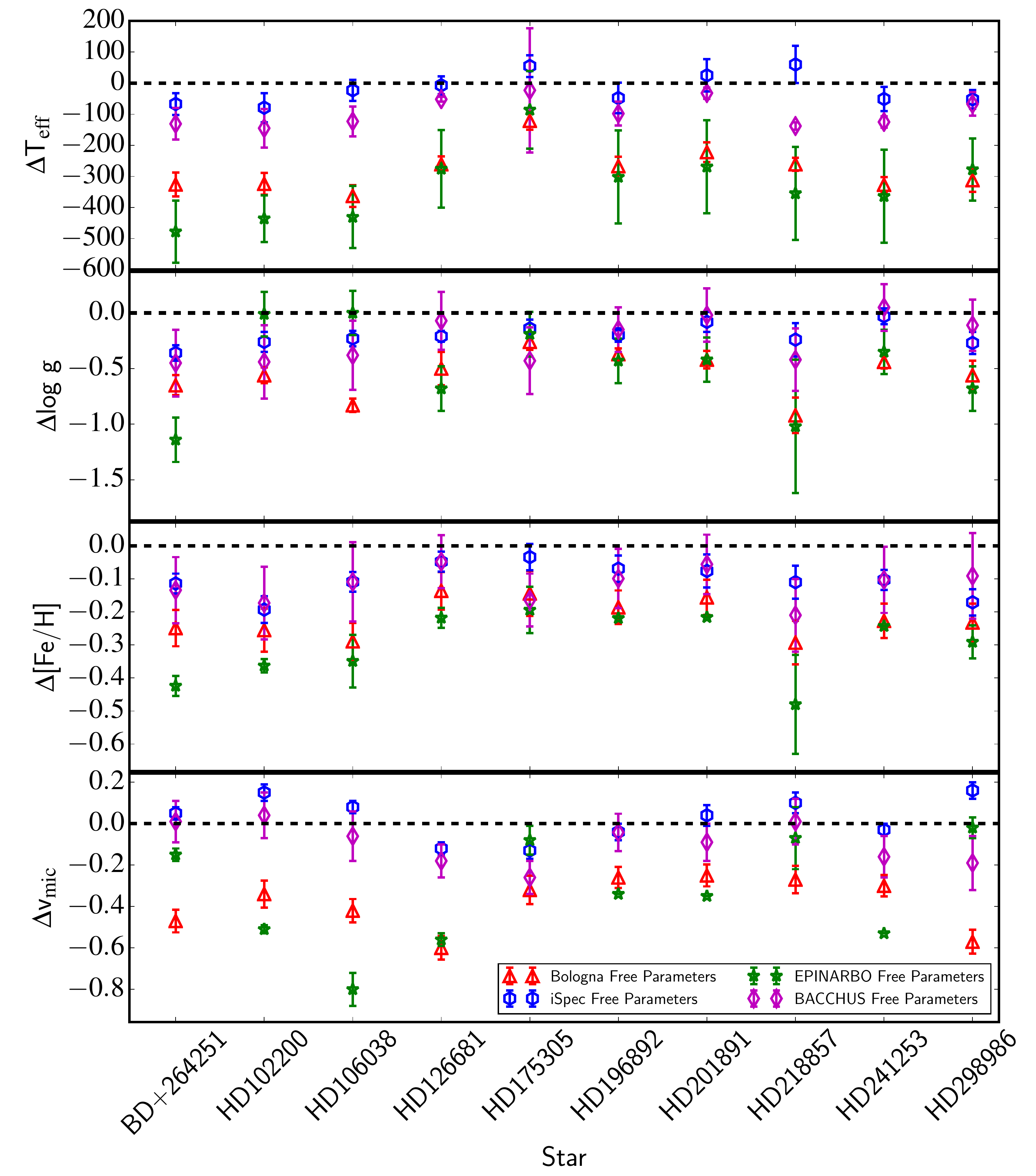}
\caption{The $\Delta$\teff, $\Delta$\logg, $\Delta$\feh, and $\Delta$\vmic\ computed from the free run (described in Sect. \ref{subsec:feh}) for each star from top to bottom, respectively. We note here that the $\Delta$ represents the difference of the node and adopted values for each parameter. For example, $\Delta$\logg~=~\logg$_{\mathrm{node}} -$~\logg$_{\mathrm{adopted}}$. The BACCHUS (ULB) method is denoted as the open magenta diamond, the iSpec method is denoted as the open blue hexagon, the EPINARBO method is denoted as a open green star and the Bologna method is denoted as an open red triangle. }
\label{fig:freerun}
\end{figure}  
In this subsection we discuss the results star-by-star. For this discussion, we remind the reader that the adopted \teff, determined via the IRFM, can be found in Col. 6 of Table \ref{tab:computedinfo}. The adopted \logg\ is determined though relating the \angdi\ and the mass. The mass is determined through isochrone fitting, using the Y$^2$ stellar evolutionary tracks, the adopted \teff\ and the mean \feh. The recommended NLTE-corrected \feh, derived using four spectroscopic methods and the adopted \teff\ and \logg, can be found in Col. 2 of Table \ref{tab:finalmet}. 

We begin the discussion by comparing the adopted \teff\ with that of the mean value from the PASTEL database and determined by the four \angdi-photometric calibrations \citep{VanBelle1999,Kervella2004,Benedetto2005,Boyajian2014}. In addition, we evaluate the spectroscopic validity of the \teff\ by ensuring that the trend in the Fe abundance with EP is null. As a diagnostic, we compare the adopted \teff\ and \teff\ from the free run (described in Sect. \ref{subsec:feh}). We note here that the results of the free run indicate that the EW methods tend to systematically underestimate the \teff\ and \logg. A potential reason for this is that the EW methods are affected by the restriction of lines allowed to be used in this analysis while synthesis methods are less affected by this. In addition, there are stark differences in the EW and synthesis procedures (e.g. sigma-clipping, convergence threshold of the pipeline, etc.) that were not fixed during this test. We stress that this test is not attempting to quantify the performance of EW methods. 

We then compare the adopted \logg\ with those determined from various means in the literature and from the free stellar parameter run. We test its validity by confirming that the \FeI\  and \FeII\ abundance agree (ionisation balance). Next we compare the metallicity derived using the adopted \teff\ and \logg\ and that from the literature. The \feh\ from the literature in most cases assumes LTE while we tabulate the NLTE corrected metallicity. The NLTE correction listed in Table \ref{tab:finalmet} is positive and thus may explain why in Fig. \ref{fig:final_lit} our final NLTE-corrected \feh\ (filled black circles) are a bit larger than the literature (open red circles). These NLTE correction are on the order of 0.05 dex. We note here that these corrections are treated as an uncertainty in our results. 

We also inspect the trend between REW and Fe abundance as a way to access the quality of the \vmic. As a general comment, the \teff\ determined using indirect data in all stars is systematically higher than the mean \teff\ from the PASTEL database (Fig. \ref{fig:final_lit}) and determined spectroscopically (Fig. \ref{fig:freerun}). We compute the combined uncertainty in the \feh\ in the same way as Paper I (i.e. by quadratically summing all $\sigma$ and $\Delta$ Cols. in Table \ref{tab:finalmet}). In addition, we remark as to whether the candidate can have direct \angdi\ measurements from current optical or near-infrared interferometers including the the VLT Interferometer or the CHARA array \citep[for a detailed description of such facilities and their \angdi\ limitation see][]{Dravins2012}. Finally, using the above discussion we either recommend or not recommend the star as a new GBS candidate.

\subsubsection*{BD+264251}
The adopted \teff\ of this star is hotter than the mean literature value by 140 K (2\%). It is most discrepant from the \teff\ derived via ($B-V$) photometry in the work of \cite{Mishenia2000}. In addition, The adopted \teff\ for this star is in fair agreement with the temperature derived from the various photometric calibration of angular diameter \citep{VanBelle1999, Kervella2004, Benedetto2005, Boyajian2014}. The \teff\ from the free run output of this star is between 0.01\% and 6\% smaller than the adopted \teff\ for the iSpec and EPINARBO nodes, respectively. The spectroscopic and adopted \teff\ do not agree which is consistent with the significant trend in the Fe abundance as a function of EP (Fig \ref{fig:nodes-abu}). However, this trend can be resolved by varying the stellar parameters within the uncertainties. In particular, it may be resolved by reducing \vmic\ by 0.2~\kms (i.e. the assumed uncertainty in the \vmic).  

The adopted \logg\ of this star is 0.1 dex (2\%) larger than the mean value from the PASTEL database. The \FeI\  and \FeII\ lines agree to within 0.05 dex (Table \ref{tab:finalmet}). In addition, the most discrepant \logg\ from the literature is from \cite{Mishenia2000}. In this study the \logg\ is derived from the ionisation balance however only making use of 20 \FeI\  and 5 \FeII\ lines. We not only make use of a method independent of spectroscopy for our adopted \logg, we also find relatively good agreement between 63 \FeI\  and 8 \FeII\ lines. 

The \feh\ derived from the spectrum assuming the adopted \teff\ and \logg\ is 0.05 dex (4\%) larger than the mean value from the PASTEL database and from 0.10 -- 0.40 dex (or 8--32\%) larger than the free run output of the ULB/BACCHUS and EPINARBO nodes, respectively. The combined uncertainty in \feh\ is on the order of $\pm$0.15 dex. There is no significant correlation between Fe abundance with REW. 

We do not recommend this star as a GBS candidate because of the discrepant photometry, ranging a total of 0.15 mag in V, which leads to relatively uncertain \teff. Additionally, the agreement between the \teff\ from the \angdi-photometric relationships and the IRFM is in worse agreement than all of the other candidates. This uncertainty in \teff\ propagates to all other parameters. In addition, the predicted \angdi\ of this star is 0.07 mas and thus will be impossible to measure directly with the current state-of-the-art interferometers (with limits on the order of 0.1 mas with the Cherenkov Telescope Array) and possibly future intensity interferometers \citep[e.g. Fig. 1 of][]{Dravins2012}. 

\subsubsection*{HD102200}
The adopted \teff\ of this star is in excellent agreement  (less than 1\%) with other spectroscopic and photometric studies \citep[e.g.][]{Mashonkina2003, Gehren2004, Jonsell2005,Sousa2011}. It is also in good agreement with the \teff\ derived from the various \angdi-photometric calibrations. We note that the \teff\ from the free run output of this star ranges between less than 0.1\% and 6\% from the adopted \teff\ for the iSpec and EPINARBO nodes, respectively. Additionally, the adopted \teff\ is consistent with the spectroscopic \teff. This is indicated by the null trend in the \feh\ abundance as a function of EP validating the adopted \teff.

The adopted \logg\ of this star is in excellent agreement (less than 0.5\%) with the mean value of studies collated in the PASTEL database. It is also in fair agreement with the free parameter run. The disagreement between the adopted value and the free parameter run ranges between 0.5 and 10\% for the EPINARBO and Bologna methods, respectively. There is also very good agreement (within 0.02 dex) between mean \FeI\  abundance, determined from averaging 58 lines and, the average \FeII\ abundance, determined by averaging 8 \FeII\ lines. This indicates that the  adopted \logg\ is in good agreement with the spectroscopic \logg. 

The \feh\ derived from the spectrum assuming the adopted \teff\ and \logg\ is $\sim$ 0.10 dex (10\%) larger than the mean literature value and 0.17 -- 0.35 dex (15--30\%) larger than the \feh\ determined in the free spectroscopic run with the ULB/BACCHUS and EPINARBO methods, respectively (Fig. \ref{fig:freerun}). However, it is important to keep in mind that both the free run and the bulk of the literature assumes LTE. The NLTE correction for this star is on the order of +0.05 dex (see Table \ref{tab:finalmet}). The combined uncertainty in \feh\ is on the order of $\pm$0.13 dex. There is no significant correlation between Fe abundance with REW.

In light of good agreement between the adopted stellar parameters and the various literature sources, the spectroscopic validation, and the free run output, we recommend this star as a GBS candidate. In addition, its predicted \angdi\ is 0.14 mas (twice as large as BD+264251). However, due to its faintness (V=8.8) it would be very challenging to achieve a direct estimate of the \angdi\ of this star with current interferometers.

\subsubsection*{HD106038}
The adopted \teff\ of this star agrees well ($\sim 2$\%) with the mean value from the literature \citep{Alonso1996, Nissen1997, Nissen2002, Ramirez2005,Gratton2003, Casagrande2011}. The most discrepant \teff\ from the literature is cooler than the adopted \teff\ by $\sim$200 K determined via ($V-K$)-\teff\ relations \citep{Nissen2002}. The adopted \teff\ is also in good agreement with the derived from the photometric calibration of \angdi\ ($\sim$1\%). Additionally, the \teff\ from the free run output of this star ranges between less than 0.1\% and 6\% from adopted \teff\ for the iSpec and EPINARBO nodes, respectively. The spectroscopic analysis showed that there is a null trend in the Fe abundance as a function EP. This indicates that the spectroscopic and adopted \teff\ are consistent with one another.

 The adopted \logg\ of this star is in good agreement (4\%)  with the mean value from the literature. It is also in good agreement the values determined from the free parameter run (between 0.5 -- 18\% for the EPINARBO and Bologna methods, respectively).  The \FeI\  and \FeII\ are consistent with each other within --0.026 dex which indicates that the adopted and spectroscopic \logg\ are in agreement. 
 
 The \feh\ derived assuming the adopted \teff\ and \logg\ is $\sim$~0.05 dex (4\%) larger than the mean from literature and 0.08 -- 0.28 dex (15--30\%) larger than the \feh\ determined from the free run from the ULB and EPINARBO methods, respectively (Fig. \ref{fig:freerun}). In addition, the combined \feh\ uncertainty is on the order of 0.13 dex. Finally, There is no significant correlation between Fe abundance with REW. 
 
We have shown that there is good agreement between the adopted stellar parameters and the various literature sources, the spectroscopic validation and the free run output. As a result we recommend this star as a GBS candidate. However, similar to BD+264251, this star has a predicted \angdi\ of 0.07 mas making it impossible to observe with current interferometers. 

\subsubsection*{HD126681}
The adopted \teff\ is in excellent agreement ($\sim 1.2$\%) with the typical \teff\ found in the literature \citep[e.g.][]{Tomkin1992, Blackwell1998, Fulbright2000, Nissen2002, Gratton2003, Reddy2006, Masana2006, Sousa2011}. The most discrepant \teff\ is from the work of \cite{Reddy2006}. The authors determine the \teff\ of their sample using Str\"{o}mgen ($b-y$) photometry \citep[e.g.][]{Alonso1996b}. However, we note that at the \teff\ of this star, the authors show (in their Fig. 6) that the difference in \teff\ determined by Str\"{o}mgen ($b-y$) photometry and ($V-K$) photometry has a dispersion of at least 100 K. The adopted \teff\ is also consistent with those derived from the photometric calibrations of \angdi\ within 100 K. The \teff\ from the free run output of this star ranges between less than 0.1\% and 4\% from adopted \teff\ for the iSpec and EPINARBO nodes, respectively. However,  there is a significant trend in the \feh\ abundance as a function of EP. This trend cannot be resolved by accounting for the uncertainties in the stellar parameters. This indicates that the adopted \teff\ is not in good agreement with the spectroscopic \teff. 

The adopted \logg\ of this star is in good agreement with the mean value from the PASTEL database (4\%). It is also in good agreement with the free run output (between 0.5 -- 18\% for the EPINARBO and Bologna methods, respectively). In addition, the mean abundance of \FeI\  (using 61 neutral lines) and \FeII\  (using 7 ionised lines) agrees within 0.021 dex. 

The \feh\ derived from the spectrum assuming the adopted \teff\ and \logg\ (described in Sect. \ref{subsec:feh}) is $\sim$ 0.05 dex (4\%) larger than the mean literature value and 0.08 -- 0.28 dex (15--30\%) larger than the \feh\ determined in the free parameter run by the ULB and EPINARBO methods, respectively (Fig. \ref{fig:freerun}). The NLTE corrections on the Fe abundance are on the order of +0.02 dex.  The combined uncertainty in the \feh\ is on the order of 0.10 dex. We also found a significant correlation between Fe abundance with REW indicating that the \vmic\ may not be adequate.

We do not recommend this star as a GBS candidate because we cannot validate its \teff\ using \FeI\ ionisation/excitation balance. In addition the \vmic\ must be changed in order to balance the correlation between Fe abundance and REW. The \angdi\ of this star is on the order of 0.10 mas which would make it out of reach for current interferometers. 

\subsubsection*{HD175305}
The adopted \teff\ is in excellent agreement ($\sim 1$\%) with the mean literature value \citep[e.g.][]{Wallerstein1979, Alonso1996, Nissen1997, Fulbright2000, Burris2000, Ishigaki2012}. While the most discrepant \teff\ in the literature, from \cite{Fulbright2000}, is more than 400 K cooler than the adopted \teff, it is an outlier among many other studies. Disregarding this outlying study, the mean difference between the adopted \teff\ and the literature is 20 K. The \angdi\ determined from the photometric calibration from \cite{VanBelle1999} is larger by nearly a factor of five compared to that of \cite{Benedetto2005}. This in turn causes the temperature to be discrepant by 250 K ($\sim$4.5\%) between these calibrations. The adopted \teff\ is consistent with \cite{Benedetto2005}. The discrepancy between these two \angdi-photometric calibrations is in part what motivated using the IRFM as the adopted procedure. The adopted \teff\ and the \teff\ derived from the free run output agrees within 2\%. There is a null trend in the \feh\ abundance as a function of EP indicating good agreement between the spectroscopic and adopted \teff. 

The adopted \logg\ of this star agrees within 1\% of the mean value from the literature and those determined from the free spectroscopic run (less than 15\%). While there is an offset of 0.08 dex between the abundance of \FeI, determined from 56 neutral Fe lines, and \FeII, determined from 8 ionised Fe lines, it can be resolved by taking into account the uncertainty in \teff\ and \logg. 

The \feh\ derived from the spectrum assuming the adopted \teff\ and \logg\ is $\sim$ 0.05 dex (4\%) larger than the mean from the PASTEL database and 0.08 -- 0.28 dex (15--30\%) larger than the \feh\ determined in the free spectroscopic run from the ULB and EPINARBO methods, respectively (Fig. \ref{fig:freerun}). The combined uncertainty in the \feh\ is on the order of 0.14 dex. There is also a null correlation between the Fe abundance with REW.  

Because of the good agreement (less than 2\% in \teff, 18\% in \logg\  and less than 15\% in \feh) between the various methods (i.e. the adopted, validation through Fe excitation/ionisation balance, free run output, and literature) of determining the stellar parameters, we recommend this star as a GBS candidate. In addition, the relatively large \angdi\ of this star (0.447 $\pm$ 0.006 mas), makes it possible to be observed in the near future with current interferometers.

\subsubsection*{HD196892}
The adopted \teff\ is in good agreement (less than 2\%) with the mean literature value \citep[e.g.][]{Axer1994, Jehin1999, Thevenin1999, Gratton2003, Jonsell2005, Sousa2011}. The most discrepant \teff\ is from the work of \cite{Axer1994} where it is derived using H$\alpha$, H$\beta, $H$\gamma$, and  H$\delta$ fitting. These authors note that there are likely systematic differences of their \teff\ with photometric values from other studies \citep[e.g.][]{Fuhrmann1994}. This may, in part, explain the discrepancy. The \teff\ derived from the various photometric calibrations of \angdi\ are consistent within 100 K of the adopted value. We note that the adopted \teff\ and that from the free run output agree within 4\%. There is a significant trend in the \feh\ abundance as a function of EP indicating that the spectroscopic and adopted \teff\ disagree. This trend cannot be resolved accounting for the uncertainties in the parameters. 

The adopted \logg\ of this star is in excellent agreement (less than 1\%) with the mean value from the PASTEL database. In addition, it  is consistent with the free run. The mean \FeI\  abundance, derived using 68 neutral Fe lines is consistent (within  0.03 dex) of the \FeII\ abundance, derived from 8 ionised Fe lines. This indicates that the spectroscopic \logg\ is consistent with the adopted value. 

 The \feh\ derived from the spectrum is $\sim$ 0.1 dex (10\%) larger than the mean literature value and as much as 0.22 dex (25\%) larger than the \feh\ determined from the free spectroscopic run (Fig. \ref{fig:freerun}). We remind the reader that this is not taking into account the NLTE correction which in this star is on the order of +0.04 dex. The combined \feh\ uncertainty is on the order of $\pm$0.08 dex. There is no significant correlation between REW and Fe abundance.
 
We do not recommend this star as a GBS candidate  because of the statistically significant trend in \FeI\  abundance and EP. In particular, this trend cannot be resolved varying the parameters within their uncertainties. In addition, the \angdi\ of this star is on the order of 0.18 $\pm$ 0.002 mas  making interferometric \angdi\ measurements very challenging. 

\subsubsection*{HD201891}
This star has an adopted temperature that is in good agreement ($\sim$1\%) with the typical value from other studies \citep[e.g.][]{Edvardsson1993, Fuhrmann1997, Israelian1998, Clementini1999, Thevenin1999, Chen2000, Zhao2000, Mishenina2001, Qui2002, Ramirez2005, Valenti2005, Reddy2008, Casagrande2011}. In fact, of the 35 studies which are listed in the PASTEL database, only 7 have \teff\ that differ by more than 100 K from our adopted value. The most discrepant \teff\ is 260 K lower \citep{Valenti2005} than the adopted \teff. It is important to note that  \cite{Valenti2005} determined the \teff\ of this star using a spectral fitting procedure. In addition, the \teff\ from \cite{Valenti2005} are well calibrated around solar \teff\ and metallicity, but get increasingly worse at low metallicities and high \teff\ \citep[e.g. see Fig 11, top panel of][]{Casagrande2011}. The 1D-LTE assumption under which the \teff\ is determined through spectroscopy may also account, in part, for the discrepancy. The adopted \teff\ is also in good agreement with the four \angdi-photometric calibrations. The adopted \teff\ and free run \teff\ of this star are in fair agreement (within 4\%). While HD201891 has a statistically significant correlation between \FeI\ abundance and EP, this correlation can effectively be resolved by varying the parameters within their uncertainty. 
 
 The adopted \logg\ is in excellent agreement with the mean value from the literature. The \logg\ is also consistent (between 0 -- 10\% level for the ULB and Bologna nodes, respectively) with the free run output. There is a slight discrepancy (at the 0.06 dex level) between the mean abundance neutral Fe (using 68 \FeI\  lines) and the mean abundance of ionised Fe (using 8 \FeII\ lines). This discrepancy can be reduced to $\sim$0.02 dex by varying the parameters within their uncertainties. 
 
 The derived \feh\ is 0.07 dex (8\%) larger than the mean literature value and as much as 0.22 dex (23\%) larger than the \feh\ from the free run output. The total NLTE correction is on the order of +0.03 dex. The combined uncertainty in \feh\ is $\pm$0.10 dex. We also find no significant correlation between REW and Fe abundance. 
 
We recommend this star as a GBS candidate. While we noted a statistically significant correlation between the \FeI\  abundance and EP, this can be resolved by taking into account the uncertainties on the parameters. In addition the discrepancy between the neutral and ionised Fe lines is also reduced to an acceptable level by accounting for the uncertainties in the parameters. Finally HD201891 has a relatively high  \angdi, with \angdi\ = 0.273 $\pm$ 0.004, for a dwarf star and thus it may be possible with current interferometers to achieve an \angdi\ estimate for this star.  

\subsubsection*{HD218857}
 The adopted \teff\ is in excellent agreement (typically less than 1\%) with the literature \citep{Axer1994, Pilachowski1996, Burris2000, Mishenina2001,Ishigaki2012}. The \teff\ derived from the photometric calibration on angular diameter from \cite{Benedetto2005} is in excellent agreement with the adopted \teff. However, the photometric calibration of \cite{VanBelle1999}, is $\sim$250 K lower than the adopted value. We note that the \teff\ from the free run output of this star ranges between less than 0.1\% and 8\% from adopted \teff\ for the iSpec and EPINARBO nodes, respectively. HD218857 also has a statistically significant correlation between \FeI\ abundance and EP lines considering both high EP and all EP \FeI\  lines indicating that the \teff\ from spectroscopic techniques may be in tension with the values determined in Sect. \ref{subsec: teff}. This trend cannot be resolved by varying the stellar parameters within the uncertainty. 
 
The adopted \logg\ of this star is $\sim$0.1 dex (4\%) larger than the typical value from the literature and as much as 1 dex larger (40\%) than the value determined from the free run. However, the mean abundance of \FeI\ , determined using 56 \FeI\  lines, is within 0.01 dex of the mean abundance of \FeII\, determined using 8 \FeII\ lines. 
 
 The derived \feh\ is 0.13 dex (8\%) larger than the mean literature value and as as much as 0.48 dex (27\%) larger than the \feh\ determined in the free spectroscopic run. The typical NLTE Fe corrections for this star are on the order of +0.06 dex. The combined uncertainty in \feh\ is on the order of 0.16 dex. In addition, we find a significant correlation between REW and Fe abundance indicating a potential issue with the \vmic.
 
 We do not recommend this star as a GBS candidate because of the significant trend in \FeI\  abundance and EP as well as the uncertain \logg. This trend cannot be resolved through varying the \teff, \logg, \feh, and \vmic\ within their uncertainties). In addition, the typical uncertainties in the parameters (particularly the uncertainty in \logg) of this star are quite large compared to the other stars. However, the star is rather faint (V=8.9) making interferometric \angdi\ measurements very challenging if not impossible.

\subsubsection*{HD241253}
The adopted \teff\ of this star is 150 K (3\%) larger than the typical literature value \citep[e.g.][]{Axer1994, diBenedetto1998, Prochaska2000, Nissen2002, Gehren2004, Mashonkina2003, Masana2006, Reddy2006, Reddy2008}. The \teff\ is most discrepant with the literature at the 350 K level \citep{Reddy2006}. As we noted above these authors determine the \teff\ of their sample using Str\"{o}mgen ($b-y$) photometry. Interestingly, these authors revise the \teff\ of the star two years later \citep{Reddy2008} which makes it consistent with our adopted \teff. The \teff\ determined using the \angdi-photometric calibration are in good agreement with the adopted value (less than 2\%). The \teff\ from the free run output of this star is in moderate agreement with the adopted \teff\ (within 7\%). However, this star has a statistically significant correlation between \FeI\  abundance and EP lines considering both high EP and all EP \FeI\  lines. This correlation cannot be resolved by varying the stellar parameters within the uncertainty. 

The adopted \logg\ of this star is 0.14 dex (2\%) less than the typical value from the literature and as much as 0.45 dex larger (10\%) than the value determined from the free run. The mean \FeI\  abundance, derived using 66 neutral Fe lines does not agree well (at the 0.10 dex level) with the \FeII\ abundance, derived from 7 ionised Fe lines. This indicates that the spectroscopic \logg\ is not consistent with the adopted value. This ionisation imbalance is not resolved taking into account the uncertainties in the parameters.

The \feh\ derived is 0.06 dex (6\%) larger than the mean literature value and up to 0.24 dex (25\%) larger than the \feh\ determined in the free spectroscopic run. The NLTE Fe corrections are on the order of +0.03 dex. There is no significant correlation between REW and Fe abundance.

We do not recommend this star as a GBS candidate because of the significant trend in \FeI\ abundance and EP as well as the disagreement between \FeI\ and \FeII. This trend cannot be resolved through varying the \teff, \logg, \feh, and \vmic\ within their uncertainties. In addition, we cannot achieve ionisation balance accounting for the uncertainties in the parameters. This star has a predicted \angdi\ that is 0.09 mas and thus is impossible to achieve with current interferometers. 

\subsubsection*{HD298986} 
The adopted \teff\ of this star is in excellent agreement ($\sim 1.5-1.7$\%) with typical values from other studies \citep[e.g.][]{Axer1994, Nissen2002, Mashonkina2003, Masana2006, Casagrande2010, Casagrande2011}. The adopted \teff\ also agrees well with those derived from the \angdi-photometric calibrations. We note that the \teff\ from the free run output of this star agrees with the adopted \teff\ within 5\%. Additionally, the adopted \teff\ is consistent with the spectroscopic \teff, as indicated by a null trend in the \feh\ abundance as a function of EP. 

The adopted \logg\ of this star is within 0.02 dex (less than 1\%) of the typical value from the literature. The uncertainty in the \logg\ is on the order of 0.19 dex. While this uncertainty is on the high end, it is not significantly larger than several current GBS including $\alpha$~Tau, $\alpha$~Cet, and $\gamma$~Sge. However, these stars are very cool giants. It is also consistent with the spectroscopic value as indicated by the agreement, on the order of 0.03 dex, of mean abundance of ionised (6 lines) and neutral iron (66 lines). 

The derived \feh\ agrees within 0.06 dex (5\%) of the mean from the PASTEL database and can be as much as 0.29 dex (23\%) larger than the \feh\ determined from the free run (Fig. \ref{fig:freerun}). The NLTE corrections for Fe are on the order of +0.05 dex. The combined uncertainty in the \feh\ is on the order of 0.13 dex. While we do find significant correlation between REW and Fe abundance, this is resolved by increasing the \vmic\ within its uncertainty. 

Given the good agreement between the adopted values determined semi-independent of spectroscopy  and other studies, as well as consistent with \FeI\ ionisation and excitation balance, we recommend this star as a GBS candidate. The predicted angular diameter of this star is 0.07 mas and is below the detection limit of current interferometers. 

A summary of the consistency checks we have outlined above can be found for each star in Table \ref{tab:stardisucssion}. 

 \begin{table}
\caption{Summary of Star-by-Star Consistency Check.} 
\begin{tabular}{c c c c c c} 
\hline\hline 
Star&\teff&\logg&\feh&\vmic&\angdi \\
\hline
*BD+264251&$\times$&\checkmark&\checkmark&\checkmark&I\\
HD102200&\checkmark&\checkmark&\checkmark&\checkmark&I\\
HD106038&\checkmark&\checkmark&\checkmark&\checkmark&I\\
*HD126681&$\times$&\checkmark&\checkmark&$\times$&I\\
HD175305&\checkmark&\checkmark&\checkmark&\checkmark&P\\
*HD196892&$\times$&\checkmark&\checkmark&\checkmark&I\\
HD201891&\checkmark&\checkmark&\checkmark&\checkmark&P\\
*HD218857&$\times$&\checkmark&\checkmark&$\times$&I\\
*HD241253&$\times$&$\times$&\checkmark&\checkmark&I\\
HD298986&\checkmark&\checkmark&\checkmark&\checkmark&I\\
\hline \hline
\end{tabular}

 \tablefoot{In this table the \checkmark\ represents a star that has `passed' (or $\times$ for `failed') a consistency check for the \teff\ (column 2),  \logg\ (column 3),  \feh\ (column 4), and \vmic (column 5) parameters. In addition, we remark whether the \angdi\ of the star is possible (P) or impossible (I) to directly measure with current (or near future) interferometers.}
\label{tab:stardisucssion}
\end{table}

\subsection{Recommendations}

From the above discussion, we recommend the following metal-poor stars as GBS candidates for calibration and validation purposes:  HD102200, HD106038, HD175305, HD201891, and HD298986. A summary of the consistency checks and discussion can be found in Table \ref{tab:stardisucssion}. The other five stars do not pass the primary criteria for good GBS candidates. In most cases, these stars are not recommended due to not being able to validate (through Fe excitation balance) the \teff\ of the star. The stars BD+264251, HD126681, HD196892, HD218857, HD241253 are denoted with an astrix in Table \ref{tab:computedinfo} and \ref{tab:finalmet} to indicate that they are not recommended as GBS candidates. 



%
%
%
%
%

\section{Summary and Conclusions} \label{sec:conclusion}
In this paper, we make an analysis of a sample of well-studied metal-poor stars in order to evaluate which of them can be included as {\it Gaia} benchmark stars. The GBS are a necessary set of calibrator stars that have already been invaluable in the era of large spectroscopic surveys. These surveys (e.g. Gaia-ESO, GALAH, and others) use them to calibrate their automated stellar parameter pipeline. As the astronomical community continues to lean towards even larger spectroscopic surveys (e.g. 4MOST and WEAVE) the need for improved samples of GBS will increase. Therefore, the aim of this paper was to add stars to the metal-poor gap defined by $-2.0 <$ \feh\ $< -1.0$ dex. We initially began with 21 stars all within the desired metallicity range, however, only 10 stars remained for spectral analysis of which 5 were ultimately recommended for calibration purposes (details on their selection and all quality control cuts can be found in Sect. \ref{sec:sample+method}). Six of the ten stars in our sample were initially suggested in Appendix B of Paper I. In this work we, performed an analysis on the stellar parameters that are consistent with the previous set of GBS.

We used up to four \angdi-photometric calibrations to estimate the \angdi\ using the broad band photometry available for each star. The bolometric fluxes were computed also using photometric calibrations. This procedure has been also employed for 6 stars (20~\%) in the current GBS (Paper I). These together were used to determine the \teff\ of each star using the adopted Stefan-Boltzmann law. The \angdi-photometric calibrations of the two giant stars in our sample produced results that disagreed at the 10\% level (leading to a \teff\ discrepancy of $\sim$300~K). As such, we also employed the IRFM to estimate the \teff. We found very good agreement of the \teff\ between the IRFM and the four \angdi-photometric calibrations. The \logg\ for the stars was computed by fitting a stellar evolutionary track (from the Y$^2$ set).

The ESO and NARVAL archival spectra were then employed to derive the \feh\ for the stars. We processed (e.g. continuum normalised, convolved to common resolution of $R = 40000$, etc.) these spectra in the same way as described in Paper II. We used a set of 131 \FeI\ and \FeII\ lines from Paper III and four separate methods (nodes) to compute the \feh. There were 2 `equivalent width' nodes (EPINARBO and Bologna) and 2 spectral synthesis codes (BACCHUS and iSpec) that were used in Paper III and IV. We employed seven separate runs per node which consisted of: a main run where the \teff, \logg, and \vmic\ were fixed to their adopted value determined from the procedures outlined in Sects. \ref{subsec: teff} and \ref{subsec: logg}, and six `error' runs which varied each of the three parameters by $\pm1\sigma$ of their uncertainties. The `error' runs were used to evaluate the impact of the uncertainties in the adopted derived stellar parameters on the \feh\ analysis.

The final combined metallicity was computed as the average of that from the four nodes. The metallicity-EP and metallicity-REW plots (shown in Fig. \ref{fig:nodes-abu} and Fig.~\ref{fig:nodes-abu2}) were used to validate the stellar parameters on the basis of the standard \FeI\ ionisation/excitation balance method. We also used Fig. \ref{fig:nodes-abu} and Fig.~\ref{fig:nodes-abu2} in our discussion of the results and the star-by-star analysis noting the consistence of the adopted and spectroscopic parameters in Sect. \ref{subsec:starbystar}. We found that five of the ten stars (HD102200, HD106038, HD175305, HD201891, and HD298986) have stellar parameters which are consistent between the photometric methods and the spectroscopic analysis. In Sect. \ref{sec:results}, we evaluate the parameters in the context of the literature. 

We present, in Table \ref{tab:finalmet}, the recommended parameters of the metal-poor GBS candidates and correspond to those which do not have an asterisk. The typical uncertainties in \teff, \logg, and \feh\ are $\pm$80 K, $\pm$0.14 dex, and $\pm$0.13 dex, respectively. While these uncertainties are marginally higher compared to the current set of FGK GBS, this is likely a result of not having a direct measurement of the \angdi. We recommend all stars with large angular diameters (particularly HD175305, HD201891, and HD102200) be included in future interferometric \angdi\ studies. In fact, HD175305 and HD201891 can, in principle, be observed with current interferometers (Table \ref{tab:stardisucssion}) and a possible extension of this work is to obtain a direct \angdi\ measurement for these two stars. Direct measurement on the \angdi\ is what will be needed to improve their accuracy so that they can take their place among stars with the highest quality parameters to calibrate the next generation of surveys. 

The recommended metal-poor candidates in this paper are dominated by stars within the metallicity range of $-1.3 <$ \feh\ $< -1.0$ dex. This is a critical metallicity regime because it is the interface of several Galactic components, such as the thick disk, the accreted halo, the inner halo and potentially the metal-poor tail of the thin disk. Furthermore, there is a lack of recommended GBS at these metallicities. With this work, we have decreased the $\sim$1~dex metallicity gap by 30\% and provided the astronomical community with these urgently needed calibration stars.


In addition, In Paper IV it was shown that a line-by-line differential approach, whereby the abundance of the star of interest is compared directly with the abundance of a reference star, to derive the metallicity yields more precise results. This could be done with Fe as well to improve the precision of the metallicity values. This was not done in the present work to remain consistent with Paper III which derived the metallicity in an absolute way. Redoing the metallicity analysis of all of the GBS in a differential framework will undoubtedly improve the precision of the derived metallicities and is planned in the near future. Therefore we stress that this work was a first step. We will soon have a new version of the PASTEL catalogue (Soubiran in prep) and more precise parallaxes from Gaia which will certainly significantly increase the number of of metal-poor candidate benchmark stars.

\begin{appendix} 
\label{app:A}
\section{Description of online table}
For clarity and reproducibility of our analysis we are providing ten online tables. There is one table per star, each of which contains the information, on a line-by-line basis, to reproduce this work. These tables have the same format and structure. Table \ref{tab:online} displays the structure of the online tables which can be found in electronic format the CDS. 
\begin{table}[h]
\caption{Online Table Format.} 
\begin{tabular}{c c c } 
\hline\hline 
%
Column &Label & Unit\\
\hline
(1) &Element & \\
(2) &Absorption line wavelength & \AA \\
(3) &Mean EW & m\AA \\
(4) &Mean Abundance (A) & dex \\
(5) &NLTE correction$^a$ & dex\\
(6) &EW (EPI) & m\AA \\
(7) &EW (BOL) & m\AA \\
(8) &EW (ULB) & m\AA \\
(9) &EW (iSpec) & m\AA \\
(10) &A(EPI) & dex \\
(11) &A(BOL) & dex \\
(12) &A(ULB) & dex \\
(13) &A(iSpec) & dex \\
\hline
\hline 
\end{tabular}
\label{tab:online}
\\ \\ \tablefoot{This table is only available in electronic form at CDS. For the EW and abundances, the node is noted in the parentheses. For example EW (EPI) denotes the EW measurement of a specific line from the EPINARBO node while A(BOL) is the log(abundance) of a specific line for the Bologna node. ($^a$) In the online table, the lines with NLTE corrections of -0.000 are those that do not have corrections available. This is done for identification purposes. In these cases, the median of the NLTE corrections of the other lines is assumed.} 
\end{table}

\end{appendix}
 
\begin{acknowledgements} 
We would like to thank the referee for constructive comments that helped improved the presentation of this work. We would like to thank A. Casey for discussions that improved this work. K.H. is funded by the British Marshall Scholarship program and the King's College, Cambridge Studentship. This work was partly supported by the European Union FP7 programme through ERC grant number 320360. U.H. acknowledges support from the Swedish National Space Board (Rymdstyrelsen).

\end{acknowledgements}

\bibliographystyle{aa} 
\bibliography{refs_paperV} 

\begin{thebibliography}{102}
\expandafter\ifx\csname natexlab\endcsname\relax\def\natexlab#1{#1}\fi

\bibitem[{{Alonso} {et~al.}(1994){Alonso}, {Arribas}, \&
  {Martinez-Roger}}]{Alonso1994}
{Alonso}, A., {Arribas}, S., \& {Martinez-Roger}, C. 1994, \aaps, 107, 365

\bibitem[{{Alonso} {et~al.}(1995){Alonso}, {Arribas}, \&
  {Martinez-Roger}}]{Alonso1995}
{Alonso}, A., {Arribas}, S., \& {Martinez-Roger}, C. 1995, \aap, 297, 197

\bibitem[{{Alonso} {et~al.}(1996{\natexlab{a}}){Alonso}, {Arribas}, \&
  {Martinez-Roger}}]{Alonso1996}
{Alonso}, A., {Arribas}, S., \& {Martinez-Roger}, C. 1996{\natexlab{a}}, \aaps,
  117, 227

\bibitem[{{Alonso} {et~al.}(1996{\natexlab{b}}){Alonso}, {Arribas}, \&
  {Martinez-Roger}}]{Alonso1996b}
{Alonso}, A., {Arribas}, S., \& {Martinez-Roger}, C. 1996{\natexlab{b}}, \aap,
  313, 873

\bibitem[{{Alvarez} \& {Plez}(1998)}]{Alvarez1998}
{Alvarez}, R. \& {Plez}, B. 1998, \aap, 330, 1109

\bibitem[{{Axer} {et~al.}(1994){Axer}, {Fuhrmann}, \& {Gehren}}]{Axer1994}
{Axer}, M., {Fuhrmann}, K., \& {Gehren}, T. 1994, \aap, 291, 895

\bibitem[{{Bailer-Jones} {et~al.}(2013){Bailer-Jones}, {Andrae}, {Arcay},
  {Astraatmadja}, {Bellas-Velidis}, {Berihuete}, {Bijaoui}, {Carri{\'o}n},
  {Dafonte}, {Damerdji}, {Dapergolas}, {de Laverny}, {Delchambre}, {Drazinos},
  {Drimmel}, {Fr{\'e}mat}, {Fustes}, {Garc{\'{\i}}a-Torres}, {Gu{\'e}d{\'e}},
  {Heiter}, {Janotto}, {Karampelas}, {Kim}, {Knude}, {Kolka}, {Kontizas},
  {Kontizas}, {Korn}, {Lanzafame}, {Lebreton}, {Lindstr{\o}m}, {Liu},
  {Livanou}, {Lobel}, {Manteiga}, {Martayan}, {Ordenovic}, {Pichon},
  {Recio-Blanco}, {Rocca-Volmerange}, {Sarro}, {Smith}, {Sordo}, {Soubiran},
  {Surdej}, {Th{\'e}venin}, {Tsalmantza}, {Vallenari}, \&
  {Zorec}}]{Bailer-Jones2013}
{Bailer-Jones}, C.~A.~L., {Andrae}, R., {Arcay}, B., {et~al.} 2013, \aap, 559,
  A74

\bibitem[{{Bensby} {et~al.}(2014){Bensby}, {Feltzing}, \& {Oey}}]{Bensby2014}
{Bensby}, T., {Feltzing}, S., \& {Oey}, M.~S. 2014, \aap, 562, A71

\bibitem[{{Bergemann} {et~al.}(2012){Bergemann}, {Lind}, {Collet}, {Magic}, \&
  {Asplund}}]{Bergemann2012}
{Bergemann}, M., {Lind}, K., {Collet}, R., {Magic}, Z., \& {Asplund}, M. 2012,
  \mnras, 427, 27

\bibitem[{{Bertelli} {et~al.}(2008){Bertelli}, {Girardi}, {Marigo}, \&
  {Nasi}}]{Bertelli2008}
{Bertelli}, G., {Girardi}, L., {Marigo}, P., \& {Nasi}, E. 2008, \aap, 484, 815

\bibitem[{{Bertelli} {et~al.}(2009){Bertelli}, {Nasi}, {Girardi}, \&
  {Marigo}}]{Bertelli2009}
{Bertelli}, G., {Nasi}, E., {Girardi}, L., \& {Marigo}, P. 2009, \aap, 508, 355

\bibitem[{{Blackwell} \& {Lynas-Gray}(1998)}]{Blackwell1998}
{Blackwell}, D.~E. \& {Lynas-Gray}, A.~E. 1998, \aaps, 129, 505

\bibitem[{{Blackwell} {et~al.}(1980){Blackwell}, {Petford}, \&
  {Shallis}}]{Blackwell1980}
{Blackwell}, D.~E., {Petford}, A.~D., \& {Shallis}, M.~J. 1980, \aap, 82, 249

\bibitem[{{Blackwell} \& {Shallis}(1977)}]{Blackwell1977}
{Blackwell}, D.~E. \& {Shallis}, M.~J. 1977, \mnras, 180, 177

\bibitem[{{Blackwell} {et~al.}(1979){Blackwell}, {Shallis}, \&
  {Selby}}]{Blackwell1979}
{Blackwell}, D.~E., {Shallis}, M.~J., \& {Selby}, M.~J. 1979, \mnras, 188, 847

\bibitem[{{Blanco-Cuaresma} {et~al.}(2014{\natexlab{a}}){Blanco-Cuaresma},
  {Soubiran}, {Heiter}, \& {Jofr{\'e}}}]{2014A&A...569A.111B}
{Blanco-Cuaresma}, S., {Soubiran}, C., {Heiter}, U., \& {Jofr{\'e}}, P.
  2014{\natexlab{a}}, \aap, 569, A111

\bibitem[{{Blanco-Cuaresma} {et~al.}(2014{\natexlab{b}}){Blanco-Cuaresma},
  {Soubiran}, {Jofr{\'e}}, \& {Heiter}}]{2014A&A...566A..98B}
{Blanco-Cuaresma}, S., {Soubiran}, C., {Jofr{\'e}}, P., \& {Heiter}, U.
  2014{\natexlab{b}}, \aap, 566, A98, Paper II

\bibitem[{{Boeche} \& {Grebel}(2015)}]{Boeche2015}
{Boeche}, C. \& {Grebel}, E.~K. 2015, ArXiv e-prints

\bibitem[{{Boyajian} {et~al.}(2014){Boyajian}, {van Belle}, \& {von
  Braun}}]{Boyajian2014}
{Boyajian}, T.~S., {van Belle}, G., \& {von Braun}, K. 2014, \aj, 147, 47

\bibitem[{{Burris} {et~al.}(2000){Burris}, {Pilachowski}, {Armandroff},
  {Sneden}, {Cowan}, \& {Roe}}]{Burris2000}
{Burris}, D.~L., {Pilachowski}, C.~A., {Armandroff}, T.~E., {et~al.} 2000,
  \apj, 544, 302

\bibitem[{{Cantat-Gaudin} {et~al.}(2014){Cantat-Gaudin}, {Donati}, {Pancino},
  {Bragaglia}, {Vallenari}, {Friel}, {Sordo}, {Jacobson}, \&
  {Magrini}}]{Cantat-Gaudin2014}
{Cantat-Gaudin}, T., {Donati}, P., {Pancino}, E., {et~al.} 2014, \aap, 562, A10

\bibitem[{{Carpenter}(2001)}]{Carpenter2001}
{Carpenter}, J.~M. 2001, \aj, 121, 2851

\bibitem[{{Casagrande} {et~al.}(2006){Casagrande}, {Portinari}, \&
  {Flynn}}]{Casagrande2006}
{Casagrande}, L., {Portinari}, L., \& {Flynn}, C. 2006, \mnras, 373, 13

\bibitem[{{Casagrande} {et~al.}(2014){Casagrande}, {Portinari}, {Glass},
  {Laney}, {Silva Aguirre}, {Datson}, {Andersen}, {Nordstr{\"o}m}, {Holmberg},
  {Flynn}, \& {Asplund}}]{Casagrande2014}
{Casagrande}, L., {Portinari}, L., {Glass}, I.~S., {et~al.} 2014, \mnras, 439,
  2060

\bibitem[{{Casagrande} {et~al.}(2010){Casagrande}, {Ram{\'{\i}}rez},
  {Mel{\'e}ndez}, {Bessell}, \& {Asplund}}]{Casagrande2010}
{Casagrande}, L., {Ram{\'{\i}}rez}, I., {Mel{\'e}ndez}, J., {Bessell}, M., \&
  {Asplund}, M. 2010, \aap, 512, A54

\bibitem[{{Casagrande} {et~al.}(2011){Casagrande}, {Sch{\"o}nrich}, {Asplund},
  {Cassisi}, {Ram{\'{\i}}rez}, {Mel{\'e}ndez}, {Bensby}, \&
  {Feltzing}}]{Casagrande2011}
{Casagrande}, L., {Sch{\"o}nrich}, R., {Asplund}, M., {et~al.} 2011, \aap, 530,
  A138

\bibitem[{{Castelli} \& {Kurucz}(2004)}]{Castelli2004}
{Castelli}, F. \& {Kurucz}, R.~L. 2004, ArXiv Astrophysics e-prints

\bibitem[{{Chen} {et~al.}(2000){Chen}, {Nissen}, {Zhao}, {Zhang}, \&
  {Benoni}}]{Chen2000}
{Chen}, Y.~Q., {Nissen}, P.~E., {Zhao}, G., {Zhang}, H.~W., \& {Benoni}, T.
  2000, \aaps, 141, 491

\bibitem[{{Clementini} {et~al.}(1999){Clementini}, {Gratton}, {Carretta}, \&
  {Sneden}}]{Clementini1999}
{Clementini}, G., {Gratton}, R.~G., {Carretta}, E., \& {Sneden}, C. 1999,
  \mnras, 302, 22

\bibitem[{{Creevey} {et~al.}(2015){Creevey}, {Th{\'e}venin}, {Berio}, {Heiter},
  {von Braun}, {Mourard}, {Bigot}, {Boyajian}, {Kervella}, {Morel}, {Pichon},
  {Chiavassa}, {Nardetto}, {Perraut}, {Meilland}, {Mc Alister}, {ten
  Brummelaar}, {Farrington}, {Sturmann}, {Sturmann}, \& {Turner}}]{Creevey2015}
{Creevey}, O.~L., {Th{\'e}venin}, F., {Berio}, P., {et~al.} 2015, \aap, 575,
  A26

\bibitem[{{Creevey} {et~al.}(2012){Creevey}, {Th{\'e}venin}, {Boyajian},
  {Kervella}, {Chiavassa}, {Bigot}, {M{\'e}rand}, {Heiter}, {Morel}, {Pichon},
  {Mc Alister}, {ten Brummelaar}, {Collet}, {van Belle}, {Coud{\'e} du
  Foresto}, {Farrington}, {Goldfinger}, {Sturmann}, {Sturmann}, \&
  {Turner}}]{Creevey2012}
{Creevey}, O.~L., {Th{\'e}venin}, F., {Boyajian}, T.~S., {et~al.} 2012, \aap,
  545, A17

\bibitem[{{Cutri} {et~al.}(2003){Cutri}, {Skrutskie}, {van Dyk}, {Beichman},
  {Carpenter}, {Chester}, {Cambresy}, {Evans}, {Fowler}, {Gizis}, {Howard},
  {Huchra}, {Jarrett}, {Kopan}, {Kirkpatrick}, {Light}, {Marsh}, {McCallon},
  {Schneider}, {Stiening}, {Sykes}, {Weinberg}, {Wheaton}, {Wheelock}, \&
  {Zacarias}}]{Cutri2003}
{Cutri}, R.~M., {Skrutskie}, M.~F., {van Dyk}, S., {et~al.} 2003, VizieR Online
  Data Catalog, 2246, 0

\bibitem[{{Dalton} {et~al.}(2014){Dalton}, {Trager}, {Abrams}, {Bonifacio},
  {L{\'o}pez Aguerri}, {Middleton}, {Benn}, {Dee}, {Say{\`e}de}, {Lewis},
  {Pragt}, {Pico}, {Walton}, {Rey}, {Allende Prieto}, {Pe{\~n}ate}, {Lhome},
  {Ag{\'o}cs}, {Alonso}, {Terrett}, {Brock}, {Gilbert}, {Ridings}, {Guinouard},
  {Verheijen}, {Tosh}, {Rogers}, {Steele}, {Stuik}, {Tromp}, {Jasko}, {Kragt},
  {Lesman}, {Mottram}, {Bates}, {Gribbin}, {Rodriguez}, {Delgado}, {Martin},
  {Cano}, {Navarro}, {Irwin}, {Lewis}, {Gonzalez Solares}, {O'Mahony},
  {Bianco}, {Zurita}, {ter Horst}, {Molinari}, {Lodi}, {Guerra}, {Vallenari},
  \& {Baruffolo}}]{Dalton2014}
{Dalton}, G., {Trager}, S., {Abrams}, D.~C., {et~al.} 2014, in \procspie, Vol.
  9147, Ground-based and Airborne Instrumentation for Astronomy V, 91470L

\bibitem[{{Datson} {et~al.}(2012){Datson}, {Flynn}, \&
  {Portinari}}]{Datson2012}
{Datson}, J., {Flynn}, C., \& {Portinari}, L. 2012, \mnras, 426, 484

\bibitem[{{Datson} {et~al.}(2014){Datson}, {Flynn}, \&
  {Portinari}}]{Datson2014}
{Datson}, J., {Flynn}, C., \& {Portinari}, L. 2014, \mnras, 439, 1028

\bibitem[{{de Jong} {et~al.}(2012){de Jong}, {Bellido-Tirado}, {Chiappini},
  {Depagne}, {Haynes}, {Johl}, {Schnurr}, {Schwope}, {Walcher}, {Dionies},
  {Haynes}, {Kelz}, {Kitaura}, {Lamer}, {Minchev}, {M{\"u}ller}, {Nuza},
  {Olaya}, {Piffl}, {Popow}, {Steinmetz}, {Ural}, {Williams}, {Winkler},
  {Wisotzki}, {Ansorge}, {Banerji}, {Gonzalez Solares}, {Irwin}, {Kennicutt},
  {King}, {McMahon}, {Koposov}, {Parry}, {Sun}, {Walton}, {Finger}, {Iwert},
  {Krumpe}, {Lizon}, {Vincenzo}, {Amans}, {Bonifacio}, {Cohen}, {Francois},
  {Jagourel}, {Mignot}, {Royer}, {Sartoretti}, {Bender}, {Grupp}, {Hess},
  {Lang-Bardl}, {Muschielok}, {B{\"o}hringer}, {Boller}, {Bongiorno}, {Brusa},
  {Dwelly}, {Merloni}, {Nandra}, {Salvato}, {Pragt}, {Navarro}, {Gerlofsma},
  {Roelfsema}, {Dalton}, {Middleton}, {Tosh}, {Boeche}, {Caffau}, {Christlieb},
  {Grebel}, {Hansen}, {Koch}, {Ludwig}, {Quirrenbach}, {Sbordone}, {Seifert},
  {Thimm}, {Trifonov}, {Helmi}, {Trager}, {Feltzing}, {Korn}, \&
  {Boland}}]{de_jong2012}
{de Jong}, R.~S., {Bellido-Tirado}, O., {Chiappini}, C., {et~al.} 2012, in
  Society of Photo-Optical Instrumentation Engineers (SPIE) Conference Series,
  Vol. 8446, Society of Photo-Optical Instrumentation Engineers (SPIE)
  Conference Series

\bibitem[{{De Pascale} {et~al.}(2014){De Pascale}, {Worley}, {de Laverny},
  {Recio-Blanco}, {Hill}, \& {Bijaoui}}]{2014arXiv1409.2258D}
{De Pascale}, M., {Worley}, C.~C., {de Laverny}, P., {et~al.} 2014, ArXiv
  e-prints

\bibitem[{{De Silva} {et~al.}(2015){De Silva}, {Freeman}, {Bland-Hawthorn},
  {Martell}, {de Boer}, {Asplund}, {Keller}, {Sharma}, {Zucker}, {Zwitter},
  {Anguiano}, {Bacigalupo}, {Bayliss}, {Beavis}, {Bergemann}, {Campbell},
  {Cannon}, {Carollo}, {Casagrande}, {Casey}, {Da Costa}, {D'Orazi}, {Dotter},
  {Duong}, {Heger}, {Ireland}, {Kafle}, {Kos}, {Lattanzio}, {Lewis}, {Lin},
  {Lind}, {Munari}, {Nataf}, {O'Toole}, {Parker}, {Reid}, {Schlesinger},
  {Sheinis}, {Simpson}, {Stello}, {Ting}, {Traven}, {Watson}, {Wittenmyer},
  {Yong}, \& {Zerjal}}]{De_silva2015}
{De Silva}, G.~M., {Freeman}, K.~C., {Bland-Hawthorn}, J., {et~al.} 2015, ArXiv
  e-prints

\bibitem[{{Dekker} {et~al.}(2000){Dekker}, {D'Odorico}, {Kaufer}, {Delabre}, \&
  {Kotzlowski}}]{Dekker2000}
{Dekker}, H., {D'Odorico}, S., {Kaufer}, A., {Delabre}, B., \& {Kotzlowski}, H.
  2000, in \procspie, Vol. 4008, Optical and IR Telescope Instrumentation and
  Detectors, ed. M.~{Iye} \& A.~F. {Moorwood}, 534--545

\bibitem[{{Demarque} {et~al.}(2004){Demarque}, {Woo}, {Kim}, \&
  {Yi}}]{Demarque2004}
{Demarque}, P., {Woo}, J.-H., {Kim}, Y.-C., \& {Yi}, S.~K. 2004, \apjs, 155,
  667

\bibitem[{{di Benedetto}(1998)}]{diBenedetto1998}
{di Benedetto}, G.~P. 1998, \aap, 339, 858

\bibitem[{{Di Benedetto}(2005)}]{Benedetto2005}
{Di Benedetto}, G.~P. 2005, \mnras, 357, 174

\bibitem[{{Dravins} {et~al.}(2012){Dravins}, {LeBohec}, {Jensen}, \&
  {Nu{\~n}ez}}]{Dravins2012}
{Dravins}, D., {LeBohec}, S., {Jensen}, H., \& {Nu{\~n}ez}, P.~D. 2012, \nar,
  56, 143

\bibitem[{{Edvardsson} {et~al.}(1993){Edvardsson}, {Andersen}, {Gustafsson},
  {Lambert}, {Nissen}, \& {Tomkin}}]{Edvardsson1993}
{Edvardsson}, B., {Andersen}, J., {Gustafsson}, B., {et~al.} 1993, \aap, 275,
  101

\bibitem[{{Eisenstein} {et~al.}(2011){Eisenstein}, {Weinberg}, {Agol},
  {Aihara}, {Allende Prieto}, {Anderson}, {Arns}, {Aubourg}, {Bailey},
  {Balbinot}, \& et~al.}]{Eisenstein2011}
{Eisenstein}, D.~J., {Weinberg}, D.~H., {Agol}, E., {et~al.} 2011, \aj, 142, 72

\bibitem[{{Fuhrmann} {et~al.}(1994){Fuhrmann}, {Axer}, \&
  {Gehren}}]{Fuhrmann1994}
{Fuhrmann}, K., {Axer}, M., \& {Gehren}, T. 1994, \aap, 285, 585

\bibitem[{{Fuhrmann} {et~al.}(1997){Fuhrmann}, {Pfeiffer}, {Frank}, {Reetz}, \&
  {Gehren}}]{Fuhrmann1997}
{Fuhrmann}, K., {Pfeiffer}, M., {Frank}, C., {Reetz}, J., \& {Gehren}, T. 1997,
  \aap, 323, 909

\bibitem[{{Fulbright}(2000)}]{Fulbright2000}
{Fulbright}, J.~P. 2000, \aj, 120, 1841

\bibitem[{{Gehren} {et~al.}(2004){Gehren}, {Liang}, {Shi}, {Zhang}, \&
  {Zhao}}]{Gehren2004}
{Gehren}, T., {Liang}, Y.~C., {Shi}, J.~R., {Zhang}, H.~W., \& {Zhao}, G. 2004,
  \aap, 413, 1045

\bibitem[{{Gilmore} {et~al.}(2012){Gilmore}, {Randich}, {Asplund}, {Binney},
  {Bonifacio}, {Drew}, {Feltzing}, {Ferguson}, {Jeffries}, {Micela},
  {Negueruela}, {Prusti}, {Rix}, {Vallenari}, {Alfaro}, {Allende-Prieto},
  {Babusiaux}, {Bensby}, {Blomme}, {Bragaglia}, {Flaccomio}, {Fran{\c c}ois},
  {Irwin}, {Koposov}, {Korn}, {Lanzafame}, {Pancino}, {Paunzen},
  {Recio-Blanco}, {Sacco}, {Smiljanic}, {Van Eck}, \& {Walton}}]{Gilmore2012}
{Gilmore}, G., {Randich}, S., {Asplund}, M., {et~al.} 2012, The Messenger, 147,
  25

\bibitem[{{Gonz{\'a}lez Hern{\'a}ndez} \&
  {Bonifacio}(2009)}]{Gonzalez-Hernandez2009}
{Gonz{\'a}lez Hern{\'a}ndez}, J.~I. \& {Bonifacio}, P. 2009, \aap, 497, 497

\bibitem[{{Gratton} {et~al.}(1996){Gratton}, {Carretta}, \&
  {Castelli}}]{Gratton1996}
{Gratton}, R.~G., {Carretta}, E., \& {Castelli}, F. 1996, \aap, 314, 191

\bibitem[{{Gratton} {et~al.}(2003){Gratton}, {Carretta}, {Claudi}, {Lucatello},
  \& {Barbieri}}]{Gratton2003}
{Gratton}, R.~G., {Carretta}, E., {Claudi}, R., {Lucatello}, S., \& {Barbieri},
  M. 2003, \aap, 404, 187

\bibitem[{{Gratton} {et~al.}(2000){Gratton}, {Sneden}, {Carretta}, \&
  {Bragaglia}}]{Gratton2000}
{Gratton}, R.~G., {Sneden}, C., {Carretta}, E., \& {Bragaglia}, A. 2000, \aap,
  354, 169

\bibitem[{{Gustafsson} {et~al.}(2008){Gustafsson}, {Edvardsson}, {Eriksson},
  {J{\o}rgensen}, {Nordlund}, \& {Plez}}]{2008A&A...486..951G}
{Gustafsson}, B., {Edvardsson}, B., {Eriksson}, K., {et~al.} 2008, \aap, 486,
  951

\bibitem[{{Hawkins} {et~al.}(2015){Hawkins}, {Jofr{\'e}}, {Masseron}, \&
  {Gilmore}}]{Hawkins2015b}
{Hawkins}, K., {Jofr{\'e}}, P., {Masseron}, T., \& {Gilmore}, G. 2015, \mnras,
  453, 758

\bibitem[{{Hawkins} {et~al.}(2016){Hawkins}, {Masseron}, {Jofre}, {Gilmore},
  {Elsworth}, \& {Hekker}}]{Hawkins2016}
{Hawkins}, K., {Masseron}, T., {Jofre}, P., {et~al.} 2016, ArXiv
  e-prints:1604.08800

\bibitem[{{Heiter} {et~al.}(2015){Heiter}, {Jofr{\'e}}, {Gustafsson}, {Korn},
  {Soubiran}, \& {Th{\'e}venin}}]{Heiter2015}
{Heiter}, U., {Jofr{\'e}}, P., {Gustafsson}, B., {et~al.} 2015, \aap, 582, A49,
  Paper I

\bibitem[{{Ishigaki} {et~al.}(2012){Ishigaki}, {Chiba}, \&
  {Aoki}}]{Ishigaki2012}
{Ishigaki}, M.~N., {Chiba}, M., \& {Aoki}, W. 2012, \apj, 753, 64

\bibitem[{{Israelian} {et~al.}(1998){Israelian}, {Garc{\'{\i}}a L{\'o}pez}, \&
  {Rebolo}}]{Israelian1998}
{Israelian}, G., {Garc{\'{\i}}a L{\'o}pez}, R.~J., \& {Rebolo}, R. 1998, \apj,
  507, 805

\bibitem[{{Jehin} {et~al.}(1999){Jehin}, {Magain}, {Neuforge}, {Noels},
  {Parmentier}, \& {Thoul}}]{Jehin1999}
{Jehin}, E., {Magain}, P., {Neuforge}, C., {et~al.} 1999, \aap, 341, 241

\bibitem[{{Jofr{\'e}} {et~al.}(2015){Jofr{\'e}}, {Heiter}, {Soubiran},
  {Blanco-Cuaresma}, {Masseron}, {Nordlander}, {Chemin}, {Worley}, {Van Eck},
  {Hourihane}, {Gilmore}, {Adibekyan}, {Bergemann}, {Cantat-Gaudin},
  {Delgado-Mena}, {Gonz{\'a}lez Hern{\'a}ndez}, {Guiglion}, {Lardo}, {de
  Laverny}, {Lind}, {Magrini}, {Mikolaitis}, {Montes}, {Pancino},
  {Recio-Blanco}, {Sordo}, {Sousa}, {Tabernero}, \& {Vallenari}}]{Jofre2015}
{Jofr{\'e}}, P., {Heiter}, U., {Soubiran}, C., {et~al.} 2015, \aap, 582, A81,
  Paper IV

\bibitem[{{Jofr{\'e}} {et~al.}(2014){Jofr{\'e}}, {Heiter}, {Soubiran},
  {Blanco-Cuaresma}, {Worley}, {Pancino}, {Cantat-Gaudin}, {Magrini},
  {Bergemann}, {Gonz{\'a}lez Hern{\'a}ndez}, {Hill}, {Lardo}, {de Laverny},
  {Lind}, {Masseron}, {Montes}, {Mucciarelli}, {Nordlander}, {Recio Blanco},
  {Sobeck}, {Sordo}, {Sousa}, {Tabernero}, {Vallenari}, \& {Van
  Eck}}]{Jofre2014}
{Jofr{\'e}}, P., {Heiter}, U., {Soubiran}, C., {et~al.} 2014, \aap, 564, A133,
  Paper III

\bibitem[{{Jonsell} {et~al.}(2005){Jonsell}, {Edvardsson}, {Gustafsson},
  {Magain}, {Nissen}, \& {Asplund}}]{Jonsell2005}
{Jonsell}, K., {Edvardsson}, B., {Gustafsson}, B., {et~al.} 2005, \aap, 440,
  321

\bibitem[{{Kervella} {et~al.}(2004){Kervella}, {Th{\'e}venin}, {Di Folco}, \&
  {S{\'e}gransan}}]{Kervella2004}
{Kervella}, P., {Th{\'e}venin}, F., {Di Folco}, E., \& {S{\'e}gransan}, D.
  2004, \aap, 426, 297

\bibitem[{{Lemasle} {et~al.}(2014){Lemasle}, {de Boer}, {Hill}, {Tolstoy},
  {Irwin}, {Jablonka}, {Venn}, {Battaglia}, {Starkenburg}, {Shetrone},
  {Letarte}, {Francois}, {Helmi}, {Primas}, {Kaufer}, \&
  {Szeifert}}]{2014arXiv1409.7703L}
{Lemasle}, B., {de Boer}, T., {Hill}, V., {et~al.} 2014, ArXiv e-prints

\bibitem[{{Lind} {et~al.}(2012){Lind}, {Bergemann}, \& {Asplund}}]{Lind2012}
{Lind}, K., {Bergemann}, M., \& {Asplund}, M. 2012, \mnras, 427, 50

\bibitem[{{Magrini} {et~al.}(2013){Magrini}, {Randich}, {Friel}, {Spina},
  {Jacobson}, {Cantat-Gaudin}, {Donati}, {Baglioni}, {Maiorca}, {Bragaglia},
  {Sordo}, \& {Vallenari}}]{2013A&A...558A..38M}
{Magrini}, L., {Randich}, S., {Friel}, E., {et~al.} 2013, \aap, 558, A38

\bibitem[{{Masana} {et~al.}(2006){Masana}, {Jordi}, \& {Ribas}}]{Masana2006}
{Masana}, E., {Jordi}, C., \& {Ribas}, I. 2006, \aap, 450, 735

\bibitem[{{Mashonkina} \& {Gehren}(2000)}]{Mashonkina2000}
{Mashonkina}, L. \& {Gehren}, T. 2000, \aap, 364, 249

\bibitem[{{Mashonkina} {et~al.}(2003){Mashonkina}, {Gehren}, {Travaglio}, \&
  {Borkova}}]{Mashonkina2003}
{Mashonkina}, L., {Gehren}, T., {Travaglio}, C., \& {Borkova}, T. 2003, \aap,
  397, 275

\bibitem[{Masseron(2006)}]{Masseron2006}
Masseron, T. 2006, PhD thesis, Observatoire de Paris, France

\bibitem[{{Mel{\'e}ndez} {et~al.}(2010){Mel{\'e}ndez}, {Casagrande},
  {Ram{\'{\i}}rez}, {Asplund}, \& {Schuster}}]{Melendez2010}
{Mel{\'e}ndez}, J., {Casagrande}, L., {Ram{\'{\i}}rez}, I., {Asplund}, M., \&
  {Schuster}, W.~J. 2010, \aap, 515, L3

\bibitem[{{Mermilliod} {et~al.}(1997){Mermilliod}, {Mermilliod}, \&
  {Hauck}}]{Mermilliod1997}
{Mermilliod}, J.-C., {Mermilliod}, M., \& {Hauck}, B. 1997, \aaps, 124, 349

\bibitem[{{Mishenina} {et~al.}(2000){Mishenina}, {Korotin}, {Klochkova}, \&
  {Panchuk}}]{Mishenia2000}
{Mishenina}, T.~V., {Korotin}, S.~A., {Klochkova}, V.~G., \& {Panchuk}, V.~E.
  2000, \aap, 353, 978

\bibitem[{{Mishenina} \& {Kovtyukh}(2001)}]{Mishenina2001}
{Mishenina}, T.~V. \& {Kovtyukh}, V.~V. 2001, \aap, 370, 951

\bibitem[{{Mucciarelli} {et~al.}(2013){Mucciarelli}, {Pancino}, {Lovisi},
  {Ferraro}, \& {Lapenna}}]{2013ApJ...766...78M}
{Mucciarelli}, A., {Pancino}, E., {Lovisi}, L., {Ferraro}, F.~R., \& {Lapenna},
  E. 2013, \apj, 766, 78

\bibitem[{{Nissen} {et~al.}(2002){Nissen}, {Primas}, {Asplund}, \&
  {Lambert}}]{Nissen2002}
{Nissen}, P.~E., {Primas}, F., {Asplund}, M., \& {Lambert}, D.~L. 2002, \aap,
  390, 235

\bibitem[{{Nissen} \& {Schuster}(1997)}]{Nissen1997}
{Nissen}, P.~E. \& {Schuster}, W.~J. 1997, \aap, 326, 751

\bibitem[{{Nissen} \& {Schuster}(2010)}]{Nissen2010}
{Nissen}, P.~E. \& {Schuster}, W.~J. 2010, \aap, 511, L10

\bibitem[{{Pilachowski} {et~al.}(1996){Pilachowski}, {Sneden}, \&
  {Kraft}}]{Pilachowski1996}
{Pilachowski}, C.~A., {Sneden}, C., \& {Kraft}, R.~P. 1996, \aj, 111, 1689

\bibitem[{{Plez}(2012)}]{Plez2012}
{Plez}, B. 2012, {Turbospectrum: Code for spectral synthesis}, astrophysics
  Source Code Library

\bibitem[{{Prochaska} {et~al.}(2000){Prochaska}, {Naumov}, {Carney},
  {McWilliam}, \& {Wolfe}}]{Prochaska2000}
{Prochaska}, J.~X., {Naumov}, S.~O., {Carney}, B.~W., {McWilliam}, A., \&
  {Wolfe}, A.~M. 2000, \aj, 120, 2513

\bibitem[{{Qui} {et~al.}(2002){Qui}, {Zhao}, {Takada-Hidai}, {Chen}, {Takeda},
  {Noguchi}, {Sadakane}, \& {Aoki}}]{Qui2002}
{Qui}, H.-M., {Zhao}, G., {Takada-Hidai}, M., {et~al.} 2002, \pasj, 54, 103

\bibitem[{{Ram{\'{\i}}rez} \& {Mel{\'e}ndez}(2005)}]{Ramirez2005}
{Ram{\'{\i}}rez}, I. \& {Mel{\'e}ndez}, J. 2005, \apj, 626, 446

\bibitem[{{Randich} {et~al.}(2013){Randich}, {Gilmore}, \& {Gaia-ESO
  Consortium}}]{Randich2013}
{Randich}, S., {Gilmore}, G., \& {Gaia-ESO Consortium}. 2013, The Messenger,
  154, 47

\bibitem[{{Reddy} \& {Lambert}(2008)}]{Reddy2008}
{Reddy}, B.~E. \& {Lambert}, D.~L. 2008, \mnras, 391, 95

\bibitem[{{Reddy} {et~al.}(2006){Reddy}, {Lambert}, \& {Allende
  Prieto}}]{Reddy2006}
{Reddy}, B.~E., {Lambert}, D.~L., \& {Allende Prieto}, C. 2006, \mnras, 367,
  1329

\bibitem[{{Sch{\"o}nrich} \& {Bergemann}(2014)}]{2014MNRAS.443..698S}
{Sch{\"o}nrich}, R. \& {Bergemann}, M. 2014, \mnras, 443, 698

\bibitem[{{Smiljanic} {et~al.}(2014){Smiljanic}, {Korn}, {Bergemann}, {Frasca},
  {Magrini}, {Masseron}, {Pancino}, {Ruchti}, {San Roman}, {Sbordone}, {Sousa},
  {Tabernero}, {Tautvai{\v s}ien{\.e}}, {Valentini}, {Weber}, {Worley},
  {Adibekyan}, {Allende Prieto}, {Barisevi{\v c}ius}, {Biazzo},
  {Blanco-Cuaresma}, {Bonifacio}, {Bragaglia}, {Caffau}, {Cantat-Gaudin},
  {Chorniy}, {de Laverny}, {Delgado-Mena}, {Donati}, {Duffau}, {Franciosini},
  {Friel}, {Geisler}, {Gonz{\'a}lez Hern{\'a}ndez}, {Gruyters}, {Guiglion},
  {Hansen}, {Heiter}, {Hill}, {Jacobson}, {Jofre}, {J{\"o}nsson}, {Lanzafame},
  {Lardo}, {Ludwig}, {Maiorca}, {Mikolaitis}, {Montes}, {Morel}, {Mucciarelli},
  {Mu{\~n}oz}, {Nordlander}, {Pasquini}, {Puzeras}, {Recio-Blanco}, {Ryde},
  {Sacco}, {Santos}, {Serenelli}, {Sordo}, {Soubiran}, {Spina}, {Steffen},
  {Vallenari}, {Van Eck}, {Villanova}, {Gilmore}, {Randich}, {Asplund},
  {Binney}, {Drew}, {Feltzing}, {Ferguson}, {Jeffries}, {Micela}, {Negueruela},
  {Prusti}, {Rix}, {Alfaro}, {Babusiaux}, {Bensby}, {Blomme}, {Flaccomio},
  {Fran{\c c}ois}, {Irwin}, {Koposov}, {Walton}, {Bayo}, {Carraro}, {Costado},
  {Damiani}, {Edvardsson}, {Hourihane}, {Jackson}, {Lewis}, {Lind}, {Marconi},
  {Martayan}, {Monaco}, {Morbidelli}, {Prisinzano}, \&
  {Zaggia}}]{Smiljanic2014}
{Smiljanic}, R., {Korn}, A.~J., {Bergemann}, M., {et~al.} 2014, \aap, 570, A122

\bibitem[{{Soubiran} {et~al.}(2010){Soubiran}, {Le Campion}, {Cayrel de
  Strobel}, \& {Caillo}}]{Soubiran2010}
{Soubiran}, C., {Le Campion}, J.-F., {Cayrel de Strobel}, G., \& {Caillo}, A.
  2010, \aap, 515, A111

\bibitem[{{Sousa} {et~al.}(2011){Sousa}, {Santos}, {Israelian}, {Lovis},
  {Mayor}, {Silva}, \& {Udry}}]{Sousa2011}
{Sousa}, S.~G., {Santos}, N.~C., {Israelian}, G., {et~al.} 2011, \aap, 526, A99

\bibitem[{{Stetson} \& {Pancino}(2008)}]{Stetson2008}
{Stetson}, P.~B. \& {Pancino}, E. 2008, \pasp, 120, 1332

\bibitem[{{Th{\'e}venin} \& {Idiart}(1999)}]{Thevenin1999}
{Th{\'e}venin}, F. \& {Idiart}, T.~P. 1999, \apj, 521, 753

\bibitem[{{Tomkin} {et~al.}(1992){Tomkin}, {Lemke}, {Lambert}, \&
  {Sneden}}]{Tomkin1992}
{Tomkin}, J., {Lemke}, M., {Lambert}, D.~L., \& {Sneden}, C. 1992, \aj, 104,
  1568

\bibitem[{{Valenti} \& {Fischer}(2005)}]{Valenti2005}
{Valenti}, J.~A. \& {Fischer}, D.~A. 2005, \apjs, 159, 141

\bibitem[{{van Belle}(1999)}]{VanBelle1999}
{van Belle}, G.~T. 1999, \pasp, 111, 1515

\bibitem[{{van Leeuwen}(2007)}]{VanLeeuwen2007}
{van Leeuwen}, F. 2007, \aap, 474, 653

\bibitem[{{Venn} {et~al.}(2004){Venn}, {Irwin}, {Shetrone}, {Tout}, {Hill}, \&
  {Tolstoy}}]{Venn2004}
{Venn}, K.~A., {Irwin}, M., {Shetrone}, M.~D., {et~al.} 2004, \aj, 128, 1177

\bibitem[{{Wallerstein} {et~al.}(1979){Wallerstein}, {Pilachowski}, {Gerend},
  {Baird}, \& {Canterna}}]{Wallerstein1979}
{Wallerstein}, G., {Pilachowski}, C., {Gerend}, D., {Baird}, S., \& {Canterna},
  R. 1979, \mnras, 186, 691

\bibitem[{{Yi} {et~al.}(2003){Yi}, {Kim}, \& {Demarque}}]{Yi2003}
{Yi}, S.~K., {Kim}, Y.-C., \& {Demarque}, P. 2003, \apjs, 144, 259

\bibitem[{{Zhao} \& {Gehren}(2000)}]{Zhao2000}
{Zhao}, G. \& {Gehren}, T. 2000, \aap, 362, 1077

\end{thebibliography}

\end{document}